\documentclass[trackchanges]{aastex7}
\usepackage{hyperref}

\newcommand\ha{H$\alpha$}
\newcommand\stwo{[S~II]}
\newcommand\degree{\rm $^{\circ}$}

\shorttitle{Mapping Star-Formation in Serpens-Aquila Rift}
\shortauthors{Rector et al.}
\graphicspath{{./}{figures/}}
\graphicspath{{./}{Cutouts/}}

\begin{document}

\title{Mapping Active Star-Formation in Serpens and the Aquila Rift}

\correspondingauthor{Travis A. Rector}
\email{tarector@alaska.edu}

\author[0000-0001-8164-653X]{T.~A. Rector}
\affil{Department of Physics and Astronomy, University of Alaska Anchorage, Anchorage, AK 99508, USA}
\email[show]{tarector@alaska.com}  

\author[0000-0002-6549-9792]{R. M. P. Kerr}
\affiliation{Dunlap Institute for Astronomy \& Astrophysics, University of Toronto, Toronto, ON M5S 3H4, Canada}
\email{ronan.kerr@utoronto.ca}

\author[0000-0001-7998-226X]{L. Prato}
\affil{Department of Physics, Northern Arizona University, Flagstaff, AZ 86011, USA}
\email{lisa.prato@nau.edu}

\author[0000-0002-7261-6504]{R.~Y. Shuping}
\affil{Space Science Institute, 4765 Walnut St, Suite B, Boulder, CO 80301, USA}
\email{rshuping@SpaceScience.org}

\author[0000-0003-4384-7220]{C. Bender}
\affiliation{Steward Observatory, University of Arizona, Tucson, AZ, 85719, USA}
\email{cbender@arizona.edu}

\author[0000-0001-9223-9091]{T.~L. Esplin}
\affiliation{Steward Observatory, University of Arizona, Tucson, AZ, 85719, USA}
\email{taranesplin@arizona.edu}

\author[0009-0002-2836-3447]{S. E. Abhilash}
\affiliation{Dunlap Institute for Astronomy \& Astrophysics, University of Toronto, Toronto, ON M5S 3H4, Canada}
\affiliation{Department of Chemical \& Physical Sciences, University of Toronto Mississauga\\ 3359 Mississauga Road, L5L 1C6 Mississauga ON, Canada}
\email{sharone.abhilash@mail.utoronto.ca}



\begin{abstract}

We report the results of a high-sensitivity survey for Herbig-Haro (HH) outflows in the Serpens-Aquila Rift using the DECam instrument on the \textit{Blanco} 4-meter telescope. We have detected 88 new HH objects, more than tripling the total known in this region. We have also identified likely progenitors for most of these outflows.  By combining HH object and progenitor locations with literature dust maps and Gaia astrometry, we find that HH objects mark five spatially distinct clouds hosting active star formation: the Serpens Molecular Cloud and LDN~673 at 400-500~pc, the more distant West and East components of the Aquila Rift at $\sim$600 and 700~pc, respectively, and a near cloud in Serpens at $\sim$250~pc. In both the Serpens Molecular Cloud and the Western Aquila Rift, HH objects broadly trace gas structures on the edges of low-density cavities surrounding known stellar populations, consistent with active star formation in feedback-driven shells. The near cloud in Serpens is not associated with any established stellar population, but its position and velocity suggest that it is part of the Local Bubble. And the HH objects there may mark the start of a new stellar generation connected to the Scutum North Association. Our Herbig-Haro objects therefore serve as a powerful indicator of star formation sites even in the absence of astrometrically characterized young stars, making them a useful tool for guiding future YSO surveys. 

\end{abstract}






\section{Introduction} \label{sec:intro}



First described in detail by \citet{Herbig_1951} and \citet{Haro_1953}, Herbig-Haro (HH) objects are compact nebulae created when supersonic ejecta from young-stellar objects (YSOs) collide with previous ejecta or the ambient interstellar medium \citep[e.g.,][]{Bo_Bally_2001}. They often appear as knots of emission aligned in collimated bipolar jets, ranging from parsec-scale giant flows such as HH 34 \citep{Devine_1997}, to curved jets in regions like NGC~1333 \citep{Bo_Bally_2001}, to precessing jets such as HH 30 \citep{Anglada_2007}, and even more esoteric outflows like HH 634 \citep{Magakian_2010, Maga_2016}. 
However their shapes are often complex, a result of the complicated environments in which they form and propagate as well as projection effects, often making it difficult to identify the direction toward the progenitor. While some outflows may contain several HH objects, many consist of one or only a few, further complicating the process of identifying the source. Fortunately, HH objects have unique characteristics that help to distinguish them from other sources.  Observationally, HH objects are readily identified in the optical by their \ha\ and \stwo\ emission, non-stellar appearance, and sometimes nearly collinear alignment along an outflow axis \citep{Bo_Bally_2001}. 

HH objects are relatively rare, with only $\sim$1000 known to exist.  This paucity flags them as important markers of ongoing protostellar accretion and consequential mass loss. They last for a short period of time -- roughly \(10^{3}\text{--}10^{5}\) years \citep{Zhang_2014} -- making them valuable tracers of active star formation. They also play a role in stellar feedback, as the outflows inject momentum into the molecular clouds-- potentially triggering subsequent waves of star formation or quenching it altogether \citep{B_Reipurth_1997, Grudic_2021}. HH objects are therefore crucial targets for understanding the feedback processes and early stellar evolution that shape star-forming regions by regulating, recycling, and dispersing material.  In this paper we present a survey of the Serpens-Aquila Rift in search of HH objects to characterize the star formation properties -- past and present -- of this remarkably extended region.

Serpens-Aquila contains numerous structures at different distances along the same set of sight lines, some of which are well-studied while others present largely unknown territory. In the Serpens region, there are three notable clouds: Serpens Main, Serpens Northeast, and Serpens South, which is often grouped with the H II region W40 \citep{Herczeg19}. 
All of these are associated with young stellar populations at similar distances. \textit{Gaia} and VLBI astrometry of Serpens~Main and Serpens~South indicate mean distances of 440-465~pc \citep{2018ApJ...869L..33O,Herczeg19,2023A&A...673L...1O}; and \citet{Herczeg19} proposed average distances of 460~pc and 465~pc, respectively, for W40 and Serpens Northeast, compared to 445~pc for Serpens Main. This range of distances is consistent with broad agreement in the literature that the clouds in Serpens are close to each other and appear to comprise an interrelated star-formation complex. Extensive coverage exists for both the gas and protostellar content in Serpens through survey work with Spitzer and Herschel \citep{Singh22, Pokhrel23}, while for W40 and Serpens Main there are  observations of both the CO and NH$_3$ gas \citep{Nakamura17, Shimoikura19, Pineda26}.

In addition, other structures are present along the same sight line as Serpens. Extinction surveys of the region indicate that there is a nearby component referred to as ``Serpens Cirrus'' in \citet{Herczeg19}, which is typically placed at a distance of $\sim$230~pc \citep{Straizys03}. To the galactic east, the dense clouds continue into Aquila, where dust maps show a complex network of numerous, heavily extinguished regions, including features like LDN~617, LDN~628, and LDN~673 \citep{Dobashi11}. LDN~673 is well-characterized and slightly closer than Serpens with a distance $\sim$400~pc, but its on-sky separation makes any direct connection to Serpens unlikely \citep{Herczeg19, Zhou22}. However, few coherent, large-scale observations exist for this region.
Some stellar populations in the direction of Aquila have been studied and may correspond to these clouds \citep{Prisinzano22, Kerr23}, but these relationships have yet to be explored. We also know little about the evolutionary state of Aquila. By finding HH objects, which are produced during a short-lived accretion stage that is part of the star formation process, we can confirm whether star formation is actively underway in a particular region. 

In this paper we provide a complete survey of HH objects in the Serpens-Aquila Rift, taking advantage of new DECam observations on the {\it Blanco} 4-meter telescope. In Section \ref{sec:obs}, we introduce our observations and search methodology, including a Zooniverse citizen science project. In Section \ref{sec:sources}, we summarize the HH objects in the region, newly and previously discovered, and connect each with a parent star-forming cloud. In Section \ref{sec:progenitors} we discuss the nature of our candidate HH object progenitors. Finally, we discuss the implications of our new HH object discoveries in Section \ref{sec:disc}, before concluding in Section \ref{sec:conclusions}.



\section{Data} \label{sec:obs}

\subsection{Survey Design and Observations}

\begin{figure}[ht]
\plotone{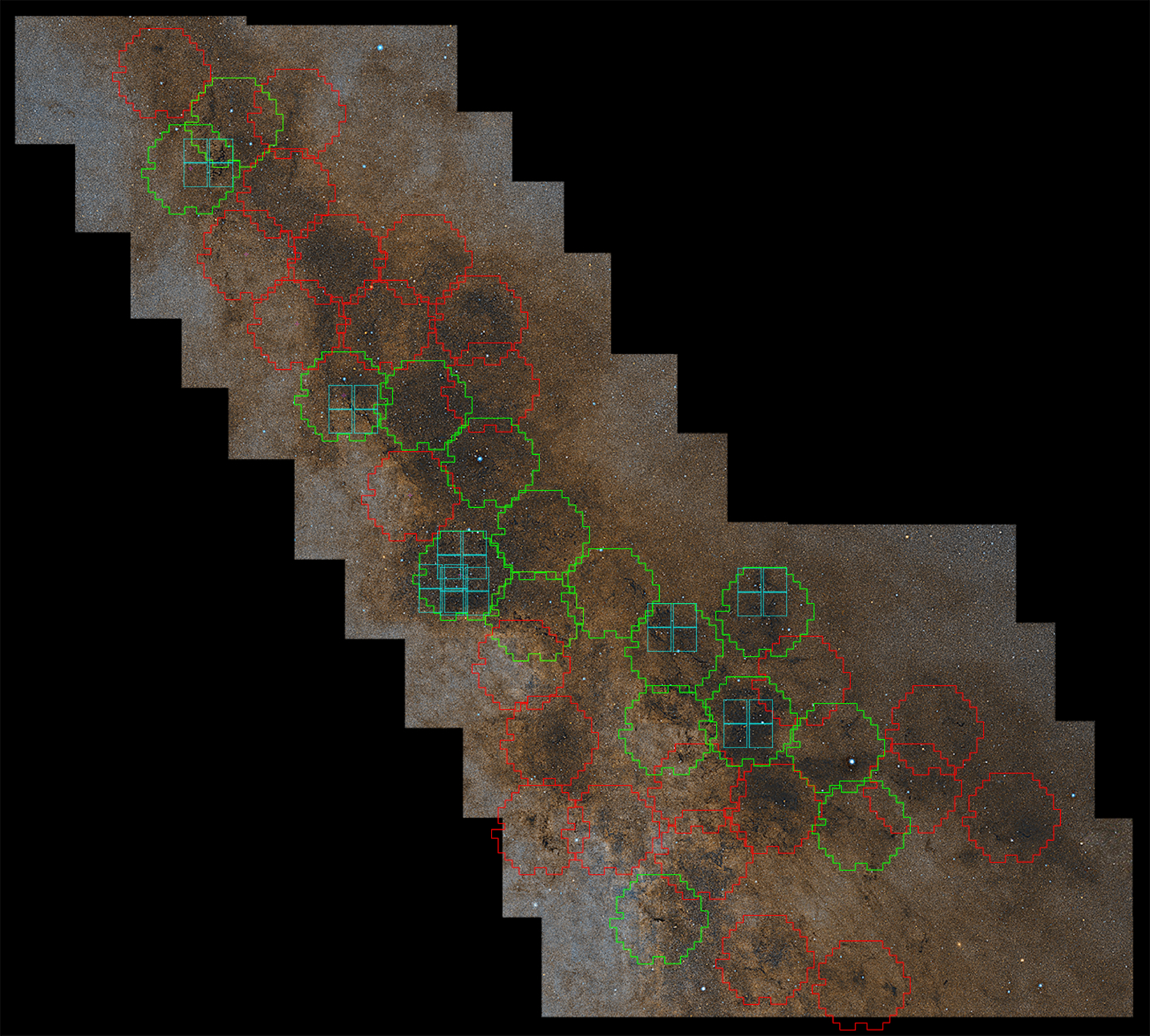}
\caption{The DECam pointings on the Serpens-Aquila cloud complex. Fields outlined in green are ones in which HH objects were found; red outlines indicate fields in which HH objects were not detected.  Bok 90Prime [SII] follow-up fields are shown in cyan.  The underlying image was obtained from PanSTARRS DR1 color ($g,r,i$) HiPS map \citep{2016arXiv161205560C} 
\label{fig:pointings}}
\end{figure}

Given the immense size of the Serpens-Aquila cloud complex, it was necessary to develop a systematic observing strategy.  Our initial study targeted  seven fields that included a range of visual extinction ($A_V$) values and small-scale structures.  A total of 21 new HH candidates were found in four of the fields, all located in or near fragmented cloud structures in ``translucent" ($1 < A_V < 5$) regions.  Based upon these results, we identified 33 additional DECam fields within the Serpens-Aquila Rift that met these two criteria.  Thus a total of 40 fields $\left( \sim120~{\rm deg}^2 \right)$ were observed, stretching from \object{LDN~673} in the northeast to below \object{W 40} and \object{Serpens South} in the southwest.  The locations of the pointings are shown in Fig.~\ref{fig:pointings}.

Observations were completed with DECam on the {\it Blanco} 4-meter telescope at the Cerro Tololo Interamerican Observatory.  DECam is a wide-field CCD imager that consists of sixty-two $2048 \times 4096$-pixel, red-sensitive CCDs, yielding 520 megapixels in total.  The field of view of a single pointing is about 3 square degrees with a scale of $0\farcs26$ pixel$^{-1}$.  

%



Observations were conducted the broadband DES {\it g,i} filters, which are similar to the SDSS survey filters of the same name.  To isolate \ha\ and \stwo\ emission, observations were also made with the DECam narrowband N662 and N673 filters-- which are centered at 6620 and 6730\AA, with FWHM values of 160 and 100\AA, respectively. Exposures were taken with a 5-pointing dither pattern to fill in chip gaps, with exposure times of $5\times240$~s in {\it g}, $5\times120$~s in {\it i}, and $5\times500$~s in N662. Observations were first taken with the {\it g}, {\it i}, and N662 filters.  For fields containing HH candidates, follow-up observations with exposure times of $5\times500$~s in the
\stwo~filter (N673) were then carried out.  DECam observations were completed between October 2020 and July 2023.  The data were reduced with the DECam Community Pipeline \citep{2014ASPC..485..379V}.  Prior to the availability of the N673 filter for DECam, follow-up observations for the first few fields were taken with the 90Prime camera on the Bok 2.3m telescope on Kitt Peak, Arizona.  The 90Prime data were collected in June 2021. The dithered images were stacked after applying flat-field and fringe correction and computing image plate solutions using Gaia DR2 astrometry using custom tools derived from IRAF.

\subsection{Analysis}

To better see faint HH objects, the {\it g, i}, and \ha\ data were combined to form color-composite images following the methodology described in \citet{2007AJ....133..598R}.  Specifically, in the color images the broadband filters $g$ and $i$ are assigned the colors cyan and yellow respectively, while \ha\ is assigned to red.  
An advantage of searching for HH objects in this manner is that the red \ha\ emission is chromatically distinct from structures visible in the other filters.  Objects that emit only in \ha\ will appear a deep red.  Furthermore, the broadband filters reveal the relative amounts of obscuration from dust and gas.  Thus, faint outflows can be found more easily in the complex environments typical of star-forming regions.  The \stwo\ observations are not included in the color-composite image but are instead used for independent confirmation. All of the HH objects are visible in the \ha\ and \stwo\ filters, but are {\it not} detected in the broadband $g$ and $i$ filters.  Thus we are confident they are sources of line emission only.

Two strategies were used to identify new HH objects in the color-composite images.  First, several team members systematically stepped through the images looking for extended red nebulosity in the color-composite images.  These objects were then confirmed in the \stwo\ grayscale images.  The searches were conducted ``double blind"; i.e., each person was unaware of the findings of the others.  

We also conducted a citizen science project on Zooniverse.org called  \href{https://www.zooniverse.org/projects/astrotrav/baby-star-search}{Baby Star Search}.  Each DECam pointing was divided into $512 \times 512$ pixel ``cutouts".  There were $\sim$2300 cutouts per pointing.  A widget was developed to blink between the color composite and \stwo\ images for each cutout, which included zoom functionality.  A tutorial was developed to train participants on how to recognize HH objects through the identification of deep red nebulosities in the color composite image {\it and} corresponding structures in the \stwo\ images.  Zooniverse investigators also learned how to identify ``imposters" such as asteroids, CR events, and charge bleeds.  
For each cutout, users were prompted to spot any candidate HH objects and to mark their locations.  
Each cutout was inspected by 15 users.  Cutouts for which three or more users indicated the presence of HH objects were inspected by the authors for confirmation.

A natural concern is that other emission-line sources could be confused as HH objects, planetary nebulae (PNe) in particular. However, they can be differentiated from HH objects by their symmetrical shapes as well as location; that is, they tend not to appear adjacent to star formation regions. We detected 30 PNe within the survey area, 3 of which are new discoveries.  Emission-line galaxies could in principle also contaminate the catalog, however the Ser-Aql Rift is at a galactic latitude of $\sim$5\degree\, resulting in very high extinction, especially behind the Serpens Clouds.  No extragalactic sources, emission-line or otherwise, are detected within the survey area.

To ensure these weren’t prior discoveries, we searched the SIMBAD \citep{2000A&AS..143....9W} and HASH \citep{2016JPhCS.728c2008P} databases. {Because HH objects often emit H$_2$ and [FeII] line emission in the infrared, we also compared the positions of our HH objects with molecular hydrogen emission-line objects (MHOs) from UWISH2 \citep{2012MNRAS.421.3257I} and extended ionized Fe objects (IFOs) from UWIFE  \citep{2024MNRAS.528.4657K}, both of which partially overlap our survey area.  This is included in our discussion of individual sources.}  The coordinates of newly discovered HH objects, and their potential progenitors, appear in Table~\ref{tbl:newHH}.  The positions correspond to the center of the brightest knot in each object.  Because the sources are extended, the coordinates are only given to a precision of 1\arcsec.  Previously discovered HH objects in the region and their progenitors are listed in Table~\ref{tbl:oldHH}.  

\begin{figure}
\gridline{\fig{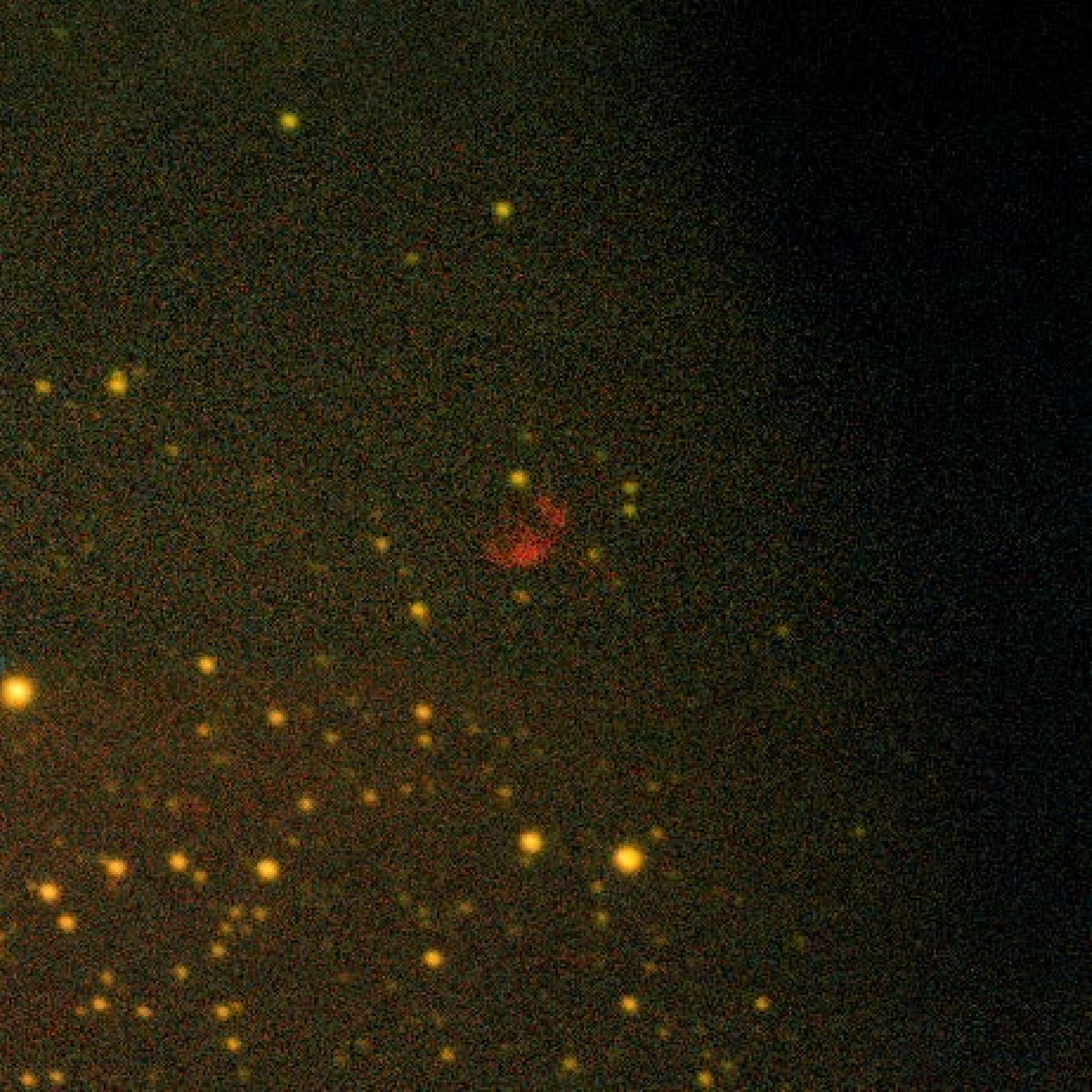}{0.33\textwidth}{HH1278}
          \fig{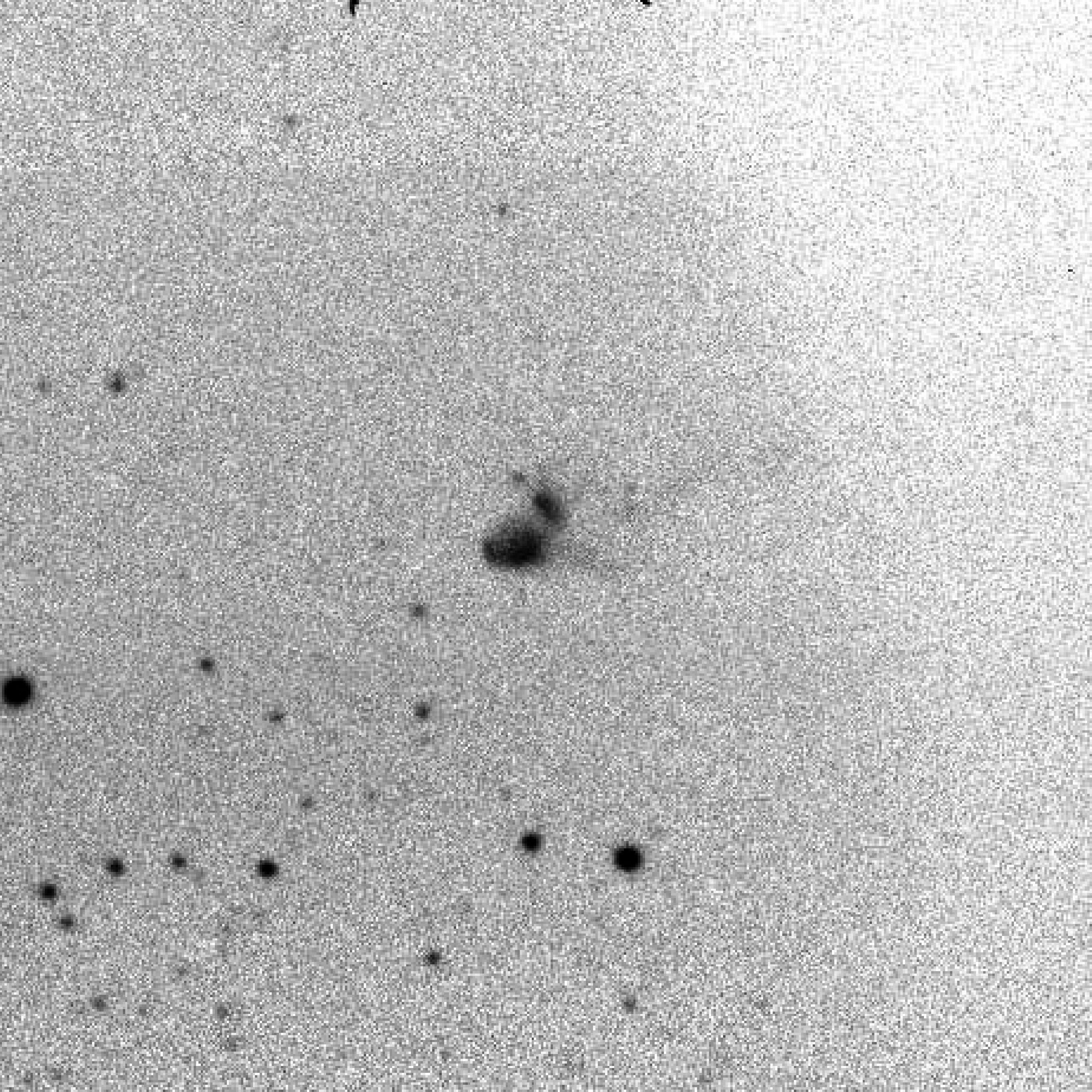}{0.33\textwidth}{HH1278}
          }
\gridline{\fig{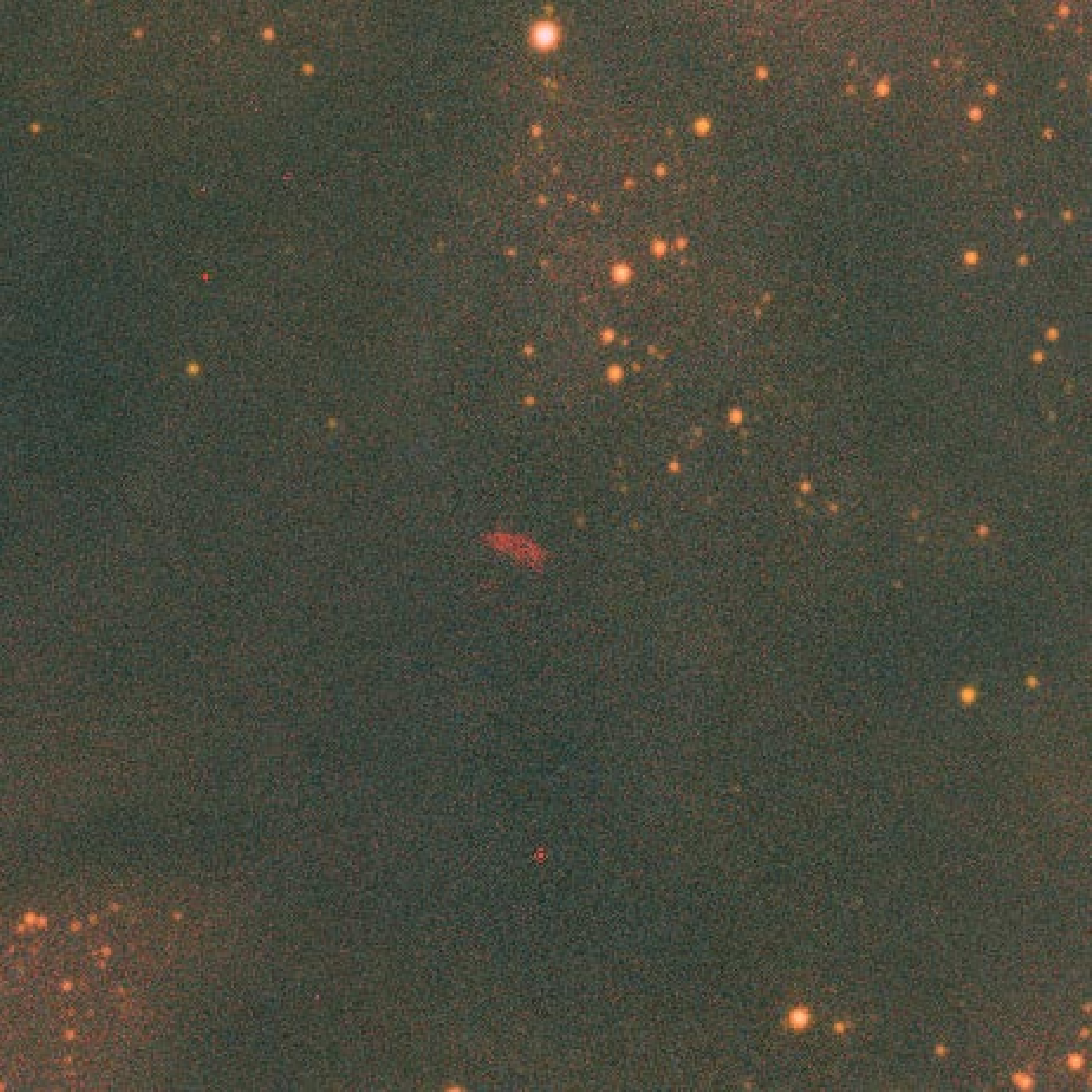}{0.33\textwidth}{HH1279}
          \fig{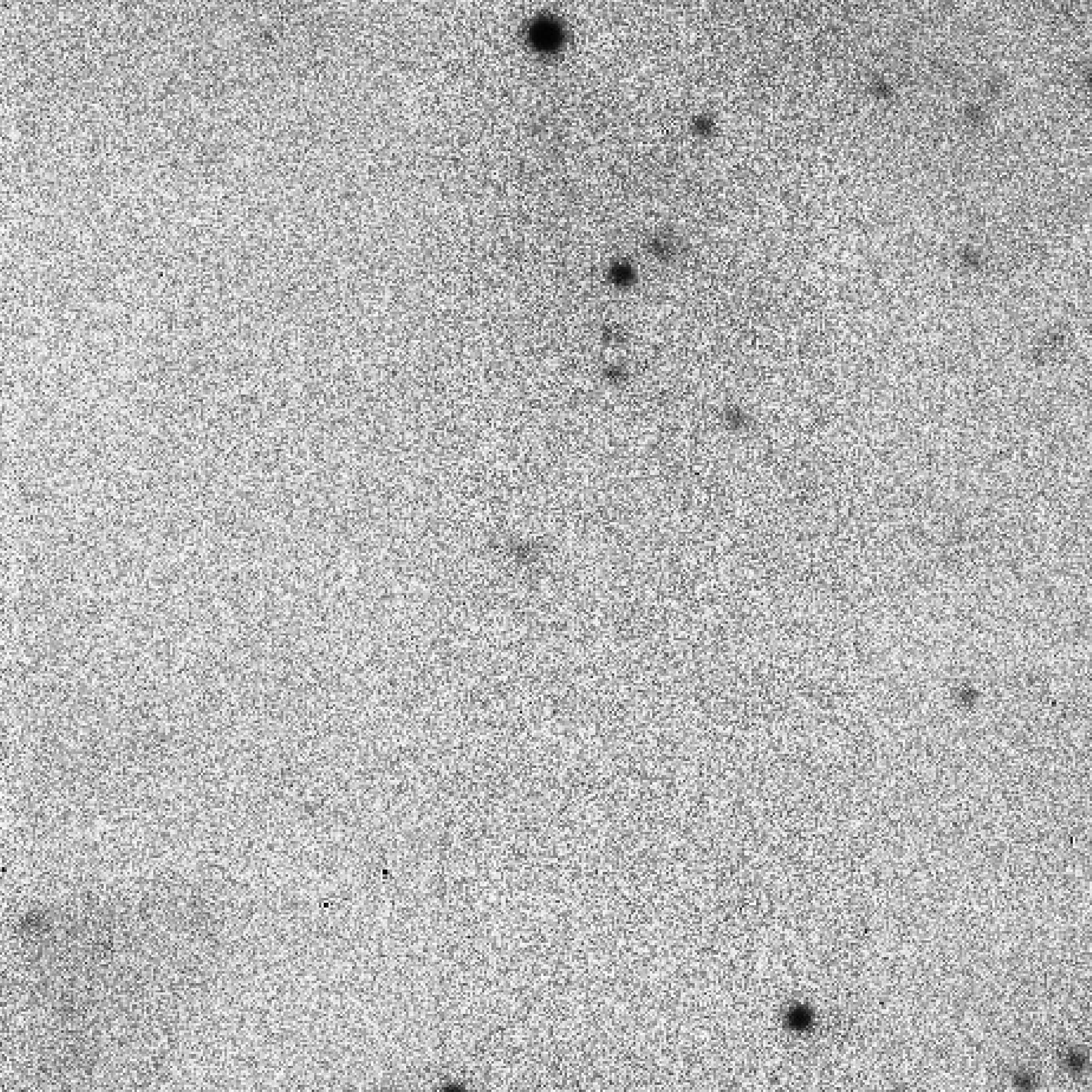}{0.33\textwidth}{HH1279}
          }
\gridline{\fig{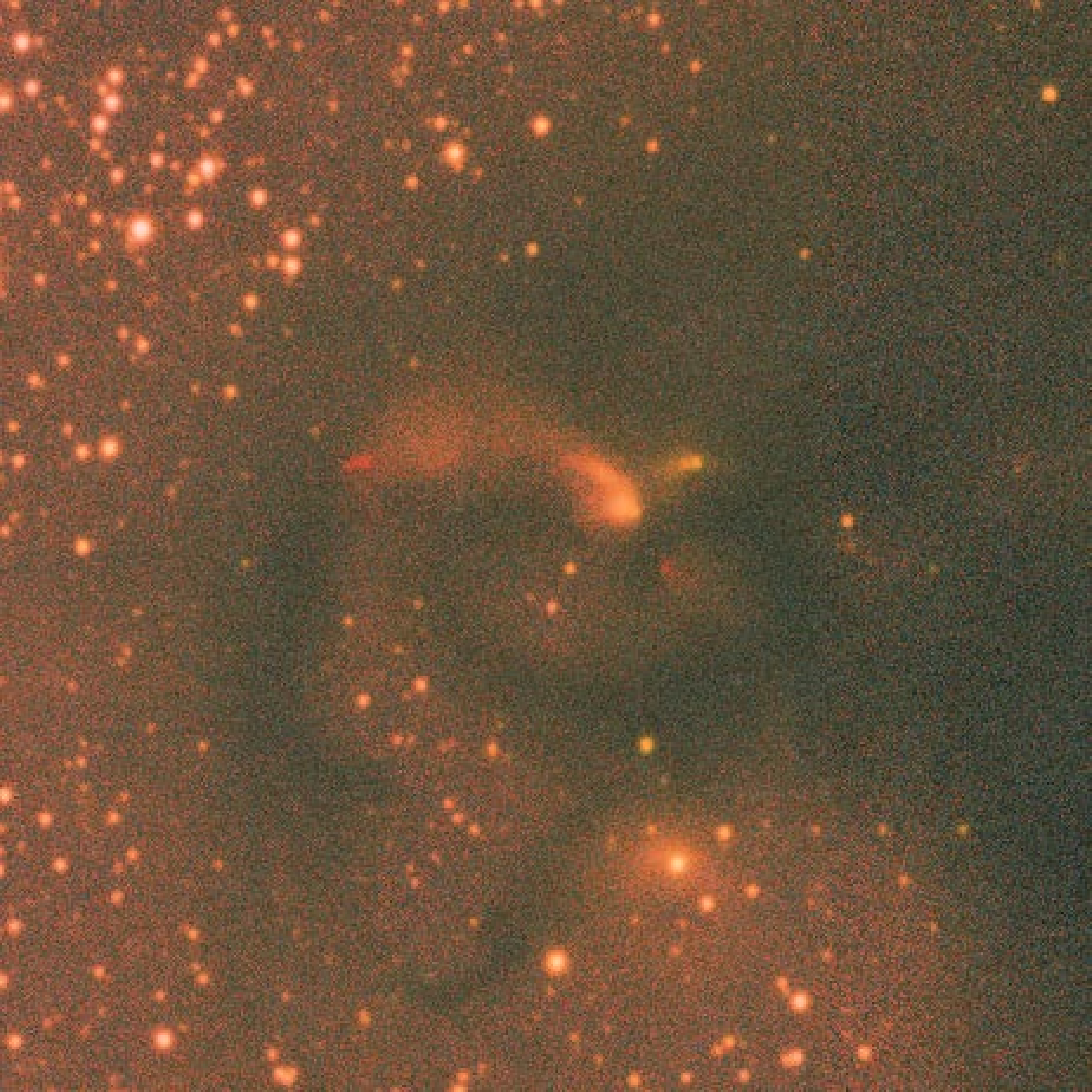}{0.33\textwidth}{HH1280 \& 1281}
          \fig{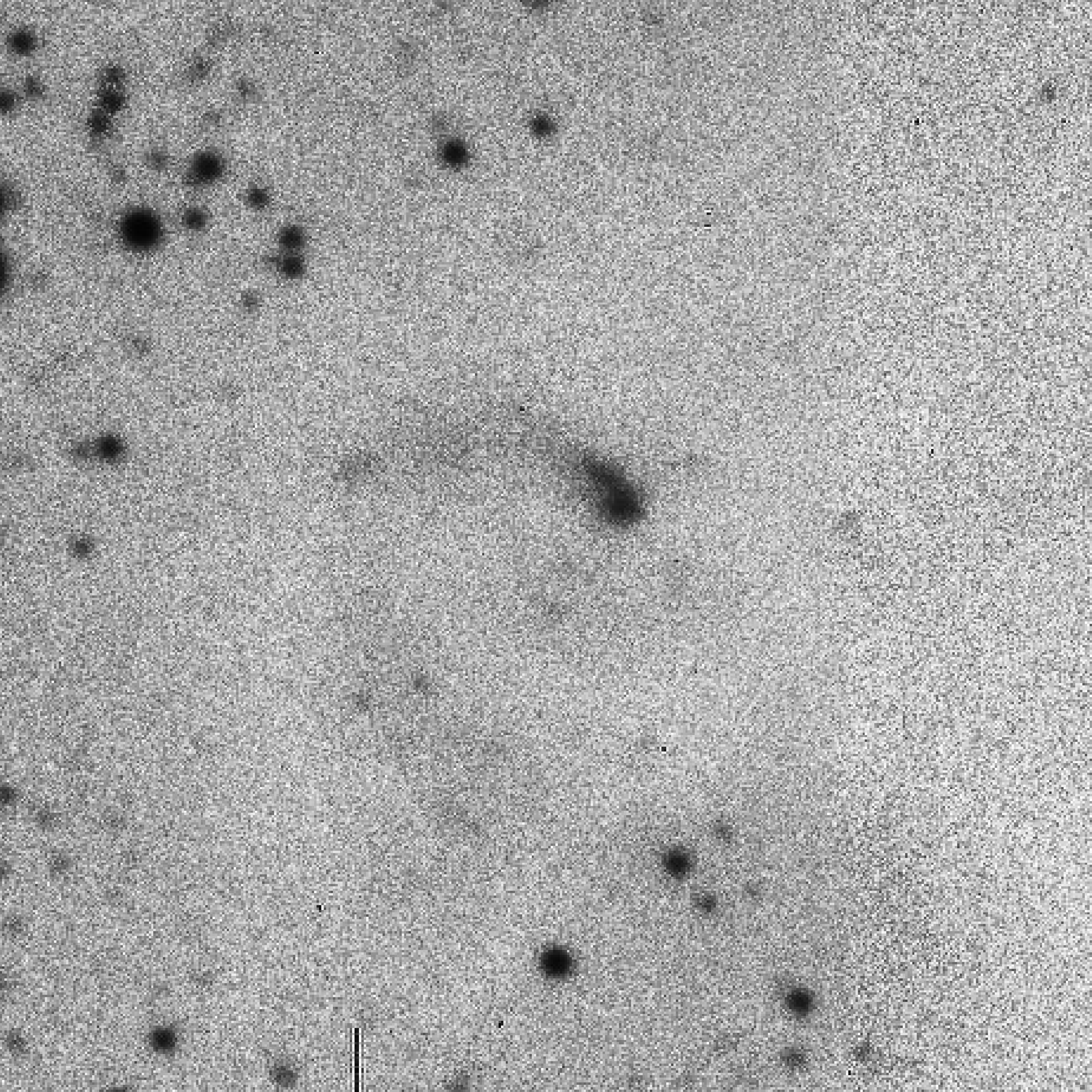}{0.33\textwidth}{HH1280 \& 1281}
          }
\caption{Example cutouts for HH1278--HH1281, newly discovered HH objects within our survey area.  Each cutout is 462 pixels (2.0\arcmin) on a side; North is up, East is to the left.  The color composite image on the left shows \ha\ emission as deep red.  The grayscale image on the right shows the \stwo\ follow-up observations.  In all cases the HH object is detected in both \ha\ and \stwo, but not in the broadband filters.  For HH1278 and HH1279 the HH object is at the center of each image; for HH1280 and HH1281, they are near the center and to the left respectively of their shared image.  The complete figure set of previously and newly discovered HH objects (192 images) is available in the online journal. 
\label{fig:hh_cutouts}}
\end{figure}

Fig.~\ref{fig:hh_cutouts} presents cutout images of all the HH objects within our survey area.  We include newly discovered individual sources from this work as well as previous discoveries. 
\stwo\ observations were completed for all sources except HH1356; we are nonetheless confident it is an HH object based upon it morphology and location.  All previously discovered HH objects within our fields were confirmed by our observations except for HH172, HH456, and HH457.  No \ha\ or \stwo\ line emission was detected at the reported coordinates of these objects  (Fig.~\ref{fig:hh_cutouts}).  Because our observations were considerably deeper, it raises doubts as to the veracity of these sources.  

\begin{deluxetable}{lllllll}
\tablecaption{Newly Discovered HH objects in Ser-Aql\label{tbl:newHH}}
\tablewidth{0pt}
\tablehead{\colhead{HH} & \colhead{RA(2000)} & \colhead{DEC} & \colhead{$l$} & \colhead{$b$} & \colhead{Progenitor candidate(s)}}
\startdata
1278 &  18:17:37.7 & -04:40:00 &      24.8944 &      5.35275 &
 IRAS 18148-0440 \\
1279  & 18:22:13.7 & -01:58:03 &      27.8314 &      5.58965 &
 IRAS 18197-0158 \\ 
1280  & 18:22:22.1 & -01:57:06 &      27.8617 &      5.56593 &
 IRAS 18197-0158 \\ 
1281  & 18:22:24.3 & -01:56:54 &      27.8689 &      5.55934 &
 IRAS 18197-0158 \\ 
1282  & 18:23:02.2 & -04:58:24 &      25.2522 &      4.01779 &
 \nodata \\
\enddata
\tablecomments{Only the first five entries of this table are shown here.  The astrometry for all of the newly discovered HH objects is published in its entirety in machine-readable format.}
\end{deluxetable}

\begin{deluxetable}{lllllll}
\tablecaption{Previously Discovered HH objects in Ser-Aql\label{tbl:oldHH}}
\tablewidth{0pt}
\tablehead{\colhead{HH} & \colhead{RA(2000)} & \colhead{DEC} & \colhead{$l$} & \colhead{$b$} & \colhead{Progenitor candidate(s)}}
\startdata
32 & 19:20:30.0 & +11:02:00 &      46.0517 &     -1.32971 & AS353A, B, or C \\
106 & 18:29:18.8 & +01:14:17 &      31.5094 &      5.48685 &
 Gaia DR3 4273733394303463936 \\
107 & 18:29:48.0 & +01:25:55 &      31.7384 &      5.46700 &
 Gaia DR3 4273733394303463936 \\
108 & 18:35:36.2 & -00:35:09 &      31.6468 &      3.79017 & IRAS 18331-0035 \\
109 & 18:35:39.0 & -00:34:20 &      31.6400 &      3.77358 & IRAS 18331-0035 \\
\enddata
\tablecomments{Only the first five entries of this table are shown here.  The astrometry for all of the previously discovered HH objects is published in its entirety in machine-readable format.}
\end{deluxetable}

\clearpage



\begin{figure}[ht]
\plotone{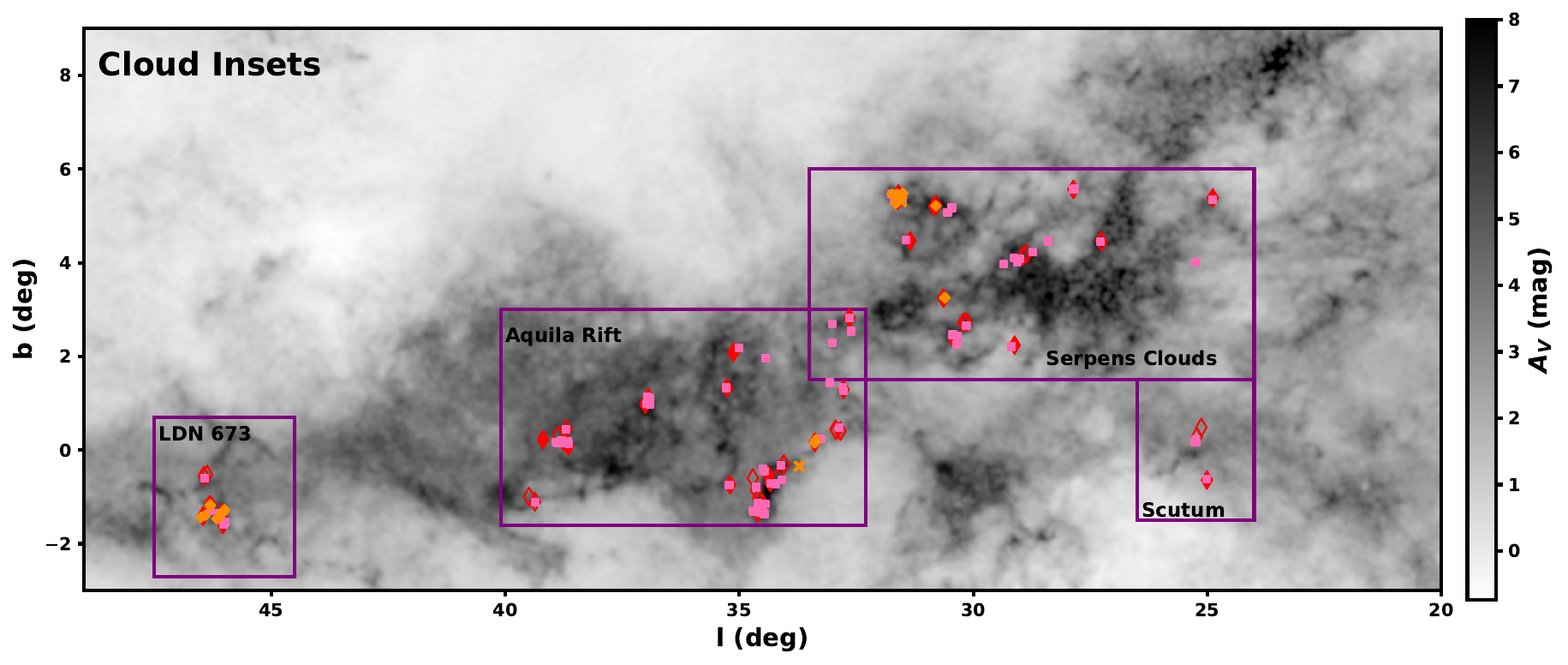}
\caption{The locations of HH objects in Serpens-Aquila. The newly discovered HH objects are marked with pink filled squares, literature HH objects are marked as wide orange diamonds (or with an orange $\times$ for HH 172, 456, and 457, which are not recovered in our work), while proposed progenitors are shown with thin red diamonds. Red open diamonds indicate cases where there are multiple possible progenitors for one HH object. Purple outlines indicate the locations of zoomed-in inset maps provided in Figs.~\ref{fig:aquilarift}, \ref{fig:serpens}, and \ref{fig:otherinsets}, highlighting sectors of the survey footprint in which HH objects are concentrated. 
\label{fig:hh_loc}}
\end{figure}

\section{Summary of Herbig-Haro Objects in the Ser-Aql Rift} \label{sec:sources}

The Serpens-Aquila Rift spans more than 20 degrees on sky, and consists of several distinct dark clouds arranged along a similar sight line, with velocities that span more than 10 km s$^{-1}$ in $v_{LSR}$ \citep{2008hsf1.book...18P}. We map the overall distribution of HH objects in our sample in Fig.~\ref{fig:hh_loc}, overlaid onto the \citet{Dobashi11} dust maps, which show extinction derived from the 2 Micron All Sky Survey (2MASS) survey on a 1$'$ grid. This figure outlines three main regions in which our HH objects reside: LDN~673 in the galactic east, the Serpens Clouds to the Northwest, and the main body of the Serpens-Aquila Rift between them. There are also two HH objects to the galactic south of the main Serpens clouds, which we provisionally refer to as the ``Scutum'' region, after the nearby constellation.  

It is essential to determine distances to the Serpens-Aquila HH objects to characterize their environments relative to ambient molecular cloud structure and feedback sources, which may also influence local star formation. We adopt \textit{Gaia} distances for the few candidate progenitors that have both a \textit{Gaia} astrometric solution and a clear connection to one of the HH objects. Elsewhere, we use \citet{Bayestar19} and \citet{Edenhofer24} dust maps to estimate the distances to the dark clouds associated with the HH objects. \citet{Edenhofer24} employ a form of iterative refinement with a correlation kernel designed to better define the positions of dust overdensities in 3d space, resulting in maps well-suited for effective distance determinations. We therefore compute distances using the \citet{Edenhofer24} dust maps where astrometric distances are not available. In more complex regions with smaller features, the $14'$ resolution of these maps may not be sufficient to resolve smaller clouds. Therefore, when there are multiple extinction peaks along a sight line in the \citet{Edenhofer24} dust maps, we use the \citet{Bayestar19} maps, which have a resolution as high as $3'.4$ per pixel, to disentangle them.

\begin{figure}[ht]
\plotone{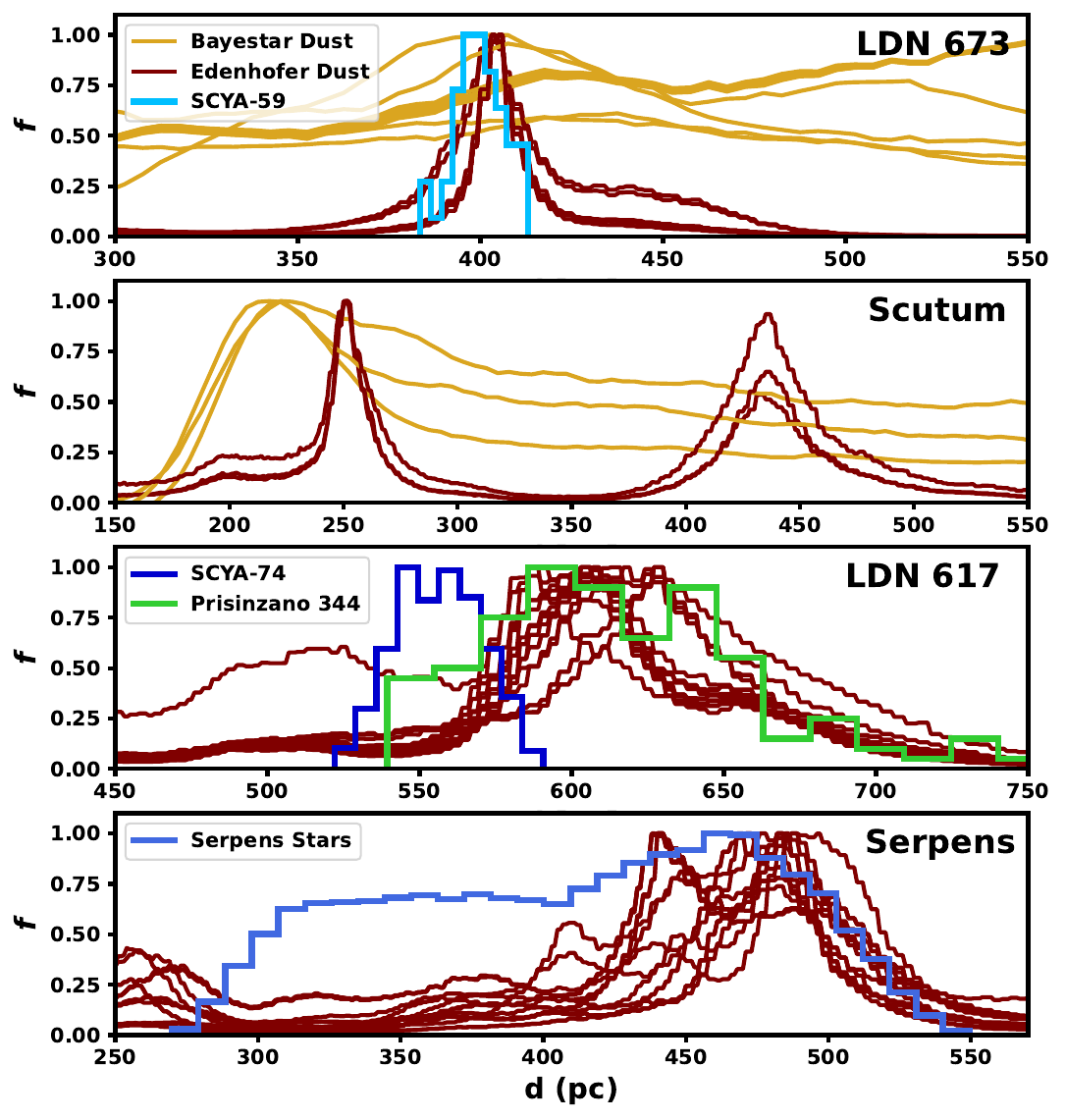}
\caption{Distance profiles of stars and dust for a selection of regions in the Serpens-Aquila Rift. In all panels, the dark red curves show extinction from the \citet{Edenhofer24} dust maps within $0.25^\circ$ of the HH object progenitor, the yellow curves show pixels within $0.1^\circ$ from the \citet{Bayestar19} dust maps, the blue step curves are histograms of distances to various populations from \citet{Kerr23} (see legends), and the green curve shows group 344 from \citet{Prisinzano22}. All curves are normalized, so the vertical axis $f$ is the fraction of the maximum value. The top panel shows LDN~673, where SCYA-59 member stars from \citet{Kerr23} stars agree closely with the \citet{Edenhofer24} dust curves, while the \citet{Bayestar19} curve roughly aligns but with a broader peak. In the second panel, we show the Scutum region, which is largely free from known stellar populations, but has multiple dust peaks from \citet{Edenhofer24} that the higher-resolution \citet{Bayestar19} maps help to disentangle. In the third panel, we show LDN~617, which aligns closely with \citet{Prisinzano22} group 344, and also overlaps with SCYA-74. Finally, in the bottom panel, we show that the dust in Serpens largely aligns with the far edge of the stellar population identified in \citet{Kerr23}. 
\label{fig:distprofs}}
\end{figure}

We show sample distance curves from both the \citet{Edenhofer24} and \citet{Bayestar19} dust maps in Fig.~\ref{fig:distprofs}. For each progenitor without a \textit{Gaia} distance, we set the distance to the peak value of the \citet{Edenhofer24} extinction curve, smoothed using a Savitsky-Golay filter with a 25~pc window length to reduce the effect of small-scale variations. The lower and upper uncertainty limits are set by computing the half width at half maximum in both directions of the peak, and converting the result to Gaussian sigma. If there are multiple peaks, we select a preferred peak using a combination of the \citet{Bayestar19} dust maps and any available \textit{Gaia} distances, as described in the analysis below. 

\subsection{Aquila Clouds}

\begin{figure}[ht]
\plotone{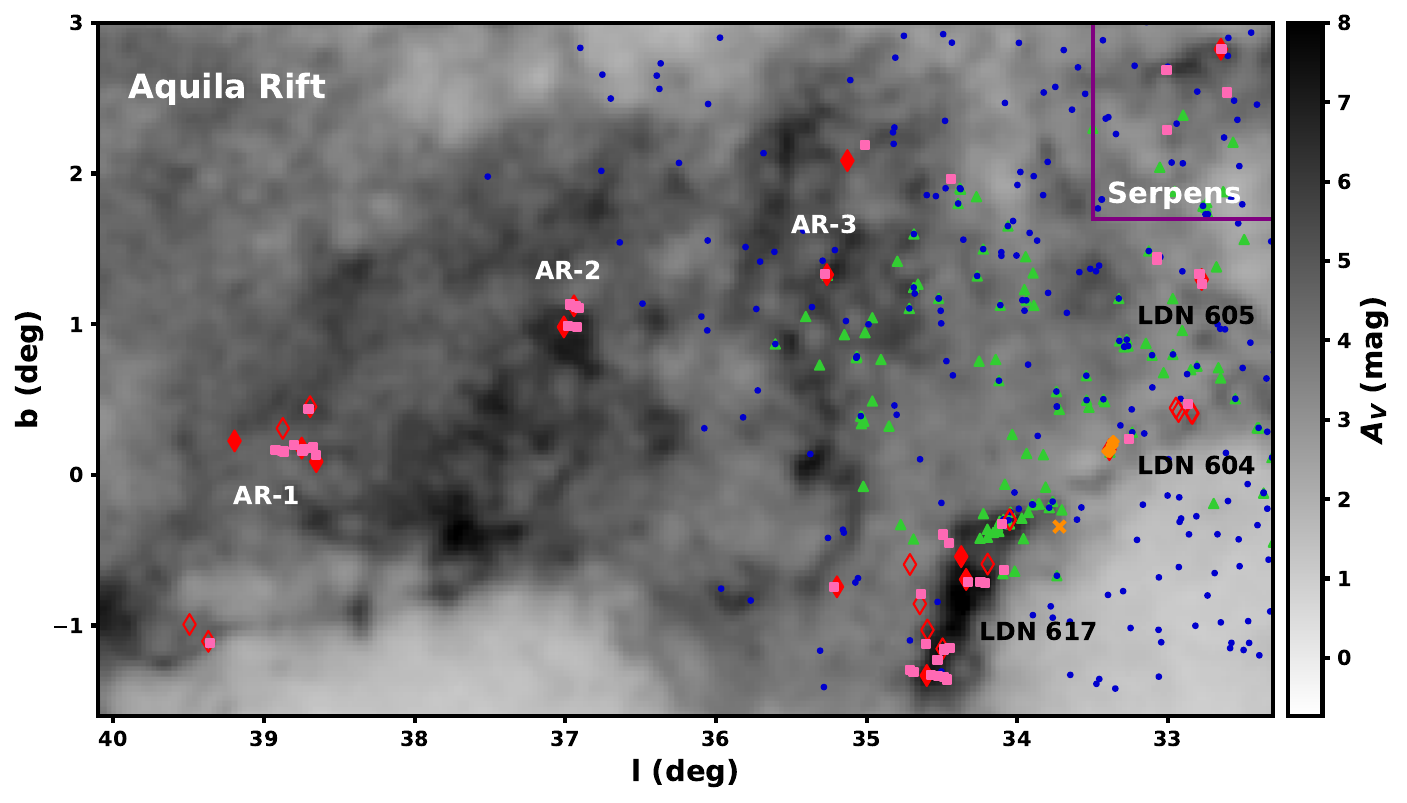}
\caption{Inset from Fig.~\ref{fig:hh_loc} of the Aquila Rift region, using the same background and markers for the HH objects and candidate progenitors. We also show stars in the SCYA-74 population from \citet{Kerr23} (blue dots) and group 344 from \citet{Prisinzano22} (green triangles), both of which overlap with the locations of the HH progenitors in 3D space, especially in the north of LDN~617. 
\label{fig:aquilarift}}
\end{figure}

\subsubsection{LDN~617}

One of the densest concentrations of HH objects in our sample is in LDN~617 and the adjoining clouds of LDN~604 and LDN~605, which form a nearly contiguous dust overdensity in the \citet{2005PASJ...57S...1D} maps, and share a similar distance in both the \citet{Bayestar19} and \citet{Edenhofer24} dust maps. In total, this region contains 31 HH objects, with 17 different potential progenitors, which we list below: 

\noindent \textbf{HH172:} This source, first reported in \citet{2002ChJAA...2...33W}, was not detected in our optical imaging (Fig.~\ref{fig:hh_cutouts}); we conclude that this is likely not an HH object. 

\noindent \textbf{HH721--724:} \citet{2007ARep...51..735N} reported the discovery of these four sources, which appear connected to LDN~604.  The progenitor(s) are embedded in the reflection nebula \object{GGD 30}.  While there may be multiple outflows, HH721, HH722, and HH723 are likely powered by the YSO \object{GGD 30 I1} \citep{2007MNRAS.375..257S}.  { IFOs 118--121 \citep{2024MNRAS.528.4657K} are co-located with these sources and are considered to be powered by the same progenitor.}

\noindent \textbf{HH1316 \& HH1319:} Both objects straddle \object{IRAS 18440+0023}, the only YSO in the dark nebula \object{Dobashi 1287} \citep{2011PASJ...63S...1D} near LDN~605. The morphology of HH1319 is also aligned with this YSO.  It is assumed to be the progenitor.

\noindent \textbf{HH1317--1318:}  Both of these objects consist of co-linear compact sources that are evenly spaced, suggesting a common source and episodic emission events.  They are seen on either side of the dark nebula \object{LDN~601}, which is adjacent to LDN~605.  While there are no known red IR sources near this location, a progenitor embedded in the cloud is the most plausible explanation.

\noindent \textbf{HH1322:}  This object lies near the dark nebula \object{SDC G32.867+0.423} and LDN~604.  There are 27 known YSOs and YSO candidates within 10\arcmin\ of this source, the nearest of which is \object{SPICY 92856}, about 4\arcmin\ to the southeast of the HH object.  Unfortunately the ambiguous morphology of the source does not indicate a preferred direction to a progenitor.

\noindent \textbf{HH1324:}  This object is superimposed on the dark nebula \object{Dobashi 1415}. The nearest YSO group is embedded in GGD~30, which contains the progenitors for HH721--724 \citep{2007MNRAS.375..257S} and is broadly connected with the LDN~604 cloud.  It is possible that HH1324 is a continuation of one of the HH721--724 outflows, although it does not appear to align with any of these objects. 

\noindent \textbf{HH1325:}  This object is embedded in the dark nebula \object{Dobashi 1430}. Located at the west end of LDN~617, this structure hosts all remaining objects listed below. The nearest YSOs are \object{MSX6C G034.0500-00.2977} and \object{MSX6C G034.0126-00.2832} \citep{2013MNRAS.430.1125C}.

\noindent \textbf{HH1331 \& HH1333:}  The proximity and morphology of these two HH objects suggest that the progenitor lies within the sub-mm source \object{HIGALBM G034.3721-00.5424}, which is embedded in the dark nebula \object{Dobashi 1442} 7\arcmin\ to the southeast of HH1333.

\noindent \textbf{HH1332:}  This object lies 7\arcmin\ to the south of the sub-mm source \object{AGAL G034.196-00.592}, which might contain the progenitor.  Alternatively, this might be an extension of the HH1334 outflow.

\noindent \textbf{HH1334:}  These three sources are part of a single outflow that aligns with the the sub-mm source \object{HIGALBM G034.3395-00.6940}, which is itself embedded in Dobashi~1442-- 5.6\arcmin to the north of HH1334B.

\noindent \textbf{HH1335:}  This object is just 1\arcmin\ to the southeast of \object{HIGALBM G034.3395-00.6940}, the source thought to power the HH1334 outflow.  This object may therefore also be part of this outflow.

\noindent \textbf{HH1336:}  There are multiple plausible progenitors in this region.  The nearest red IR source is \object{IRAS 18550+0104}, 4\arcmin\ to the east.  Alternative candidates are the YSO \object{IRAS 18555+0056}, 15\arcmin\ to the southeast, and the dense core \object{IRAS 18542+0114}, 12\arcmin\ to the northwest.  

\noindent \textbf{HH1337:}  This object is adjacent to the massive star cluster \object{G35.2-0.74N}; one of its members is likely the progenitor. { IFO~132 \citep{2024MNRAS.528.4657K} is located about 4.9\arcmin\ to the SW.}

\noindent \textbf{HH1338--1341:}  These four sources are relatively close to each other; however, the geometry requires more than one progenitor.  If HH1338, HH1339, and HH1341 are part of the same outflow, \object{IRAS 18566+0048} could conceivably be the origin, although this is unlikely.  HH1338--1340 are better aligned, and there is considerable dark nebulosity between HH1339 and HH1340.  \object{IRAS 18558+0047} is in a suitable location, and is only 42\arcsec\ from HH1339, but there are many other sources that are also plausible.

\noindent \textbf{HH1342--1347:}  This long chain of HH objects is aligned with \object{IRAS 18566+0048}, which is embedded in the dark nebula \object{DSH J1859.2+0050}.  If this is indeed a single outflow, it spans about 15\arcmin\ in total.  { IFO~135 \citep{2024MNRAS.528.4657K} is located about 6.3\arcmin\ NE of HH1346 and could also be part of this outflow.}

We show the HH objects and their progenitors in this region in the bottom-right of Fig.~\ref{fig:aquilarift}, where the three main clouds are labeled in black. Most proposed progenitors have positions that align closely with the visible dust filaments, with HH objects usually offset outward from the densest regions, especially in the core of LDN~617. This effect is also seen in several other stellar populations, such as Circinus West \citep{2025AJ....169..313R}, consistent with HH objects being observed as they break out from their parent cloud. LDN~617 is generally not associated with any well-established young stellar population. However its northern edge toward LDN~604, which is not as dense in the \citet{2005PASJ...57S...1D} dust maps, is home to overdensities of young stars as seen by both \citet{Kerr23} (SCYA-74) and \citet{Prisinzano22} (group 344). 

Two potential progenitors in this region have \textit{Gaia} astrometry, but both measurements have little constraining power with $\pi/\sigma_{\pi}<6$, so we defer to the \citet{Edenhofer24} dust maps to determine distance. The reddening peaks in these dust maps indicate an average distance to the cloud hosting these progenitors of 611~pc, with a 9~pc dispersion. This aligns closely with the stars in the region, especially as defined in \citet{Prisinzano22}, where stars in group 344 have a distance $615 \pm 45$~pc. The \citet{Edenhofer24} dust maps suggest a slight distance gradient across the region, with more distant components at $\sim$620~pc in the galactic north toward the Serpens Clouds, and distances closer to 600~pc in the main LDN~617 cloud. Because the stars from \citet{Prisinzano22}, which are more distant on average, are offset to the north, this is consistent with the stellar distribution in the region.

HH1337 is offset from LDN~617, suggesting that it may not share the same origin. It lies less than two arcminutes from the core of the dense cloud \object{G35.2-0.74N}, and within its $^{13}$CO contours from Galactic Ring Survey data \citep{Jackson06}. It is also on the edge of a peak in \object{APEX 870} $\mu$m continuum emission. \citet{Dewangan17} and \citet{WangWang23} show knots of H$_2$ emanating from the cloud, consistent with outflows, one of which matches the site of HH1337. We therefore conclude that this source is associated with G35.2-0.74N, which would place it at a distance of $2.19^{+0.24}_{-0.20}$ kpc \citep{Zhang09}, significantly further away compared to the other sources in our sample.

\subsubsection{Eastern Aquila Rift}

The remaining component of the Aquila rift to the east of LDN 617 contains several components layered along the line of sight, with some distances matching those of LDN~673 and LDN~617, and others that are much further away. The $^{12}$CO measurements from \citet{2008hsf1.book...18P} also show several velocity components, with higher $v_{LSR}$ radial velocities in clouds to the galactic east. The 18 HH objects here can be arranged into roughly three main regions, labeled in white in Fig.~\ref{fig:aquilarift}. Some small dust overdensities detected in \citet{2005PASJ...57S...1D} contain several HH objects and their progenitors, especially in the central area near LDN~628. However, the clouds in this region lack any clear unifying filamentary structure connecting the overdensities as seen in LDN~617. We summarize the HH objects found in this region below.

\noindent \textbf{HH1320:}  This object is to the west of the dark nebula \object{Dobashi 1443}, however there are no known YSOs or red IR sources embedded within this nebula, or within 30\arcmin\ of this source.

\noindent \textbf{HH1321:}  There are no dark nebulae in this region.  This object is likely associated with \object{IRAS 18454+0250}, a Herbig Ae/Be Star about 9\arcmin\ to the northeast. 

\noindent \textbf{HH1323:}  This object is embedded in the dark nebula \object{Dobashi 1497}.  Its proximity and morphology suggests that \object{IRAS 18484+0237} is the progenitor.

\noindent \textbf{HH1326--1328:}  These three HH objects are compact, colinear, and aligned with \object{HIGALBM G036.9422+01.1219}, a sub-mm source just 35\arcsec\ to the north that is embedded in a dense and compact globule.  This source is barely detected at 22\micron\ (W4) by AllWISE \citep{2010AJ....140.1868W}.

\noindent \textbf{HH1329--1330:}  These two objects are aligned with the YSO \object{IRAS 18528+0400}, which is 1.5\arcmin\ to the northeast of HH1330.  It is assumed to be the progenitor. 

\noindent \textbf{HH1348:}  This source lies along the eastern edge of the dark nebula \object{Dobashi 1616}.  Embedded in this nebula about 1\arcmin\ to the west is the sub-mm source \object{HIGALBM G038.6940+00.4527}, which is a possible progenitor although little is known about it.  Alternatively, the YSO \object{MSX6C G038.8750+00.3088} is 13\arcmin\ to the ENE.

\noindent \textbf{HH1349--1350:}  \citet{2014MNRAS.443.3388S} identified HH1350 as the planetary nebula candidate \object{PN G038.6+00.1}; however, it is a strong \stwo\ source and its morphology is more consistent with an HH object.  The alignment and morphology of HH1349--HH1350 suggest that \object{IRAS 18591+0503}, located 2.7\arcmin\ to the southeast, is the closest progenitor for both.  Alternatively, the YSO candidate SPICY~96721, located 6.0\arcmin\ to the southwest, is a plausible source.

\noindent \textbf{HH1351-1353:}  These objects lie near the dark nebula \object{Dobashi 1625}, which contains the YSO candidate \object{IRAS 18589+0511}.  HH1351 and HH1353 lie along the southern edge of the nebula and are likely associated with a source within.  HH1352 is to the north of the nebula and may instead be associated with a different progenitor.

\noindent \textbf{HH1354--1355:}  HH1354 has a distinct bow shock morphology; HH1354A is most likely a bright component of this structure.  HH1355 is to the north and is aligned with them.  The morphologies and locations of these three sources suggest that the YSO \object{CORNISH G039.1956+00.2255}, located 20\arcmin\ to the north of HH1354, is the progenitor.

\noindent \textbf{HH1356:}  This object lies directly below the unusual Orion variable \object{HBC 684} \citep{2006ApJ...647..432R} and is likely associated with it.  Alternatively, it may be associated with the YSO \object{IRAS 19045+0518}, located 11\arcmin\ to the north.  We note that \stwo\ observations were not obtained for this source; however, considering its location and morphology we are confident it is an HH object. { IFO~141, also thought to be powered by  IRAS~19045+0518, is located about 10.8\arcmin\ to the NW \citep{2024MNRAS.528.4657K}.}


The AR-1 region shown in Fig.~\ref{fig:aquilarift} is the Aquila component with the best-characterized dust distances, with clear peaks in the \citet{Edenhofer24} maps between 686 and 690~pc that broadly agree with the \citet{Bayestar19} maps. However, the candidate progenitors for HH1356 have more ambiguous distances, as the \citet{Edenhofer24} maps suggest that a far closer component at $\sim$200~pc may dominate here. \citet{Bayestar19} also shows a reddening peak at this distance, but reveals that the near component is spatially broad, and may therefore relate to distributed material near the edge of the Local Bubble \citep{ONeill24}. The dust overdensity that this source appears to be embedded in (Fig.~\ref{fig:aquilarift}) is only clearly visible between 650 and 750~pc, suggesting that this HH object is related to the same cloud as the rest of the HH progenitors in this region. 

The central component (labeled as AR-2) also has an ambiguous distance, with a bimodal distance peak between that of the progenitor candidates in AR-1 and distances around 550~pc. One of the two progenitors in this region has a \textit{Gaia} entry with a distance of 570$^{+38}_{-34}$~pc \citep{2021AJ....161..147B}, supporting a location in the closer of the two distance peaks. \citet{Bayestar19} also indicates an origin in the near component, with its higher resolution demonstrating that while a distant component is present in this region-- and may dominate given the coarser resolution of \citet{Edenhofer24}-- the area directly adjoining the clump containing these HH objects is closer. We therefore adopt the \textit{Gaia} distance for that progenitor, and use the near peak from \citet{Edenhofer24} for the second object that lacks an astrometric distance. 

The final component in this region, marked as AR-3, contains two HH objects and their progenitors in a cloud to the galactic north of LDN~617. The southern HH object is associated with a dust overdensity \citep{2005PASJ...57S...1D} that surrounds the presumed progenitor of IRAS 18484+0237, while the northern object, IRAS 18454+0250, is in a more disorganized part of the cloud. Both have broad extinction distributions in distance, indicating that there may be several clouds in close proximity along this line of sight. However, there appears to be only one major dust peak, a conclusion broadly supported by \citet{Bayestar19}. Both sources therefore appear to be at $d\sim$600~pc, but with substantial uncertainties of $\sim$60~pc on average given the broad extinction peak. IRAS 18454+0250 has a \textit{Gaia} entry, with a distance of $1456^{+61}_{-46}$ \citep{2021AJ....161..147B}. This is far more distant than the clouds we connect with this source, indicating that an association with a more distant cloud is possible. However, it is unclear whether an HH object connected with such a distant source along the galactic plane would be visible, so we favor the closer distance solution.

\subsection{Serpens Clouds} \label{sec:hhobjs-serpens}

\begin{figure}[ht]
\plotone{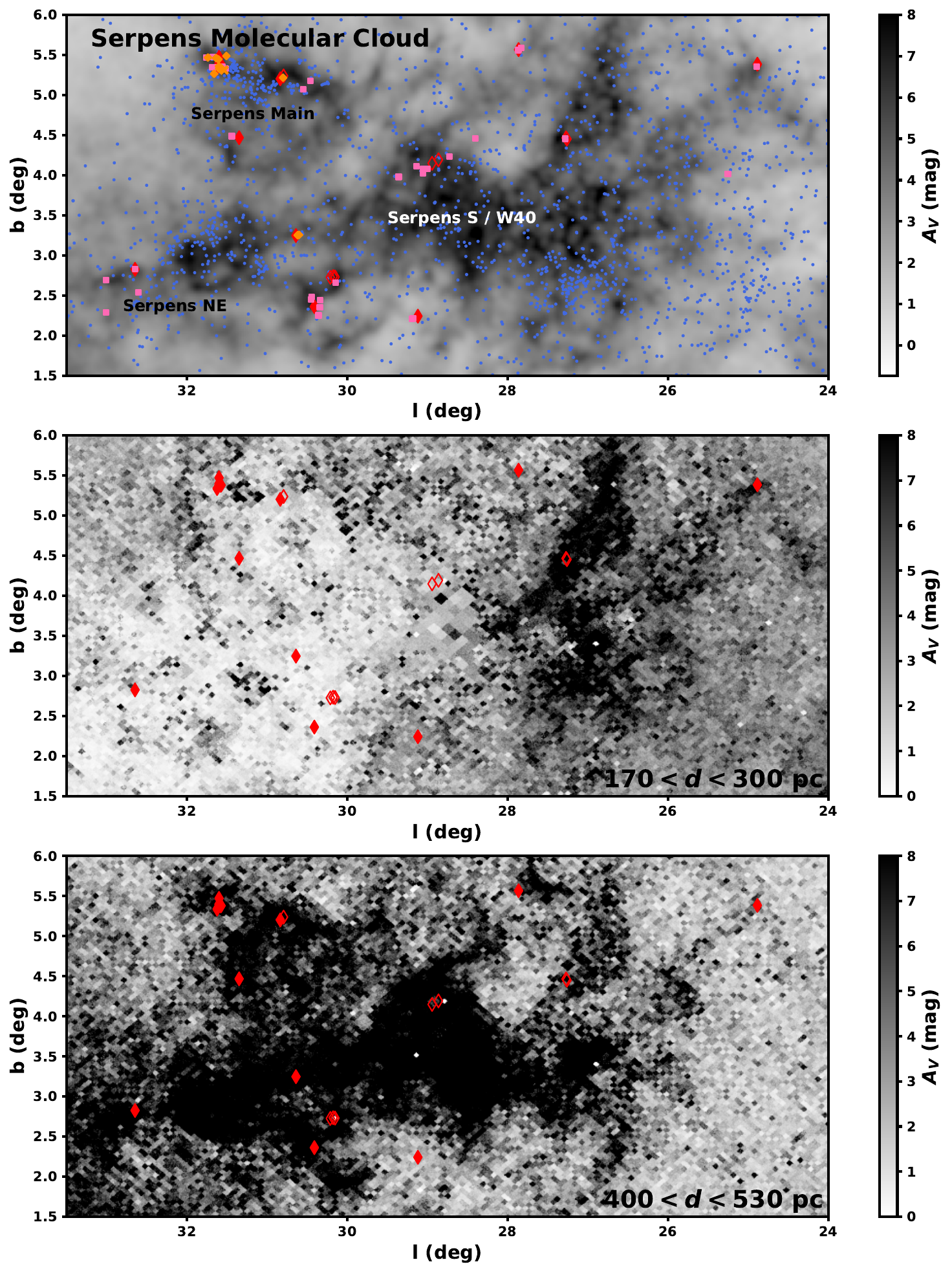}
\caption{The top panel uses the same background and markers as Fig.~\ref{fig:aquilarift}, but focuses on the Serpens region. Light blue dots mark stellar members of the Serpens association from \citet{Kerr23}. The bottom two panels show the integrated extinction across two distance ranges in the \citet{Bayestar19} maps against the progenitor candidates, showing that two clouds share this line of sight: a closer one at around 250~pc that contains the westernmost HH objects, and a more distant one connected to the main Serpens clouds at around 450~pc. 
\label{fig:serpens}}
\end{figure}

The Serpens Molecular Cloud complex is one of the largest and densest molecular clouds within 500~pc of the sun, and the region contains 52 HH objects originating from 25 potential progenitors. This complex is highly inhomogeneous, containing three different widely recognized components: Serpens South-- which corresponds to the active and highly-embedded star-forming region W40, Serpens Main, and Serpens Northeast, which are all at similar distances. A lesser-studied nearer component also exists, largely comprising the Serpens Cirrus region described in \citet{Herczeg19}. At $d\sim$250~pc, this close component likely has little relationship to the Serpens Main region of the cloud at $\sim$450~pc. 

\subsubsection{Serpens Near}

The western components of the Serpens complex, which we refer to as ``Serpens-Near," contain three HH objects, two of which can be confidently connected to unique progenitors. These objects are as follows: 

\noindent \textbf{HH1278:}  This source appears along the eastern edge of \object{LDN~483}.  It is 1.7\arcmin\ to the east and aligned with the CO outflow from the binary protostar \object{IRAS 18148-0440} \citep{2000A&A...359..967T,2022ApJ...932...34C}, which must be the progenitor.

\noindent \textbf{HH1282:}  The progenitor of this source is unknown, and no YSOs lie within 1\degree. The \object{MWC 297} cluster is 1.6\degree\ to the northeast; however, most of the dust in this area is connected to Serpens Near, not the main Serpens component where this cluster is found. An origin from a potentially undiscovered protostar in Serpens Near is therefore possible.

\noindent \textbf{HH1283:}  This object is about 50\arcsec\ and 80\arcsec\ away from the YSOs \object{2MASS J18251133-0258532} and \object{2MASS J18251332-0259549}, respectively, and both are plausible progenitors. MWC~297 and the W40 cluster, also candidates for the outflows' origin, are 1.0\degree\ and 1.8\degree\ to the southeast and northeast, respectively. The dominance of the near dust component here favors a closer origin. 


We show the distance split in Serpens in Fig.~\ref{fig:serpens}, demonstrating the alignment of these progenitors with the near components of these clouds as outlined in \citet{Bayestar19}. Our fits to the \citet{Edenhofer24} dust curves place both \object{IRAS 18148-0440} and the two potential progenitors of HH1283 at similar distances with the former, which is associated with LDN~483, being the closer of the two at $\sim$241~pc compared to $\sim$255~pc for the latter. These distances align closely with Serpens Cirrus from \citet{Herczeg19}, which has some overlap with these sources, but is largely located to their galactic west. Regardless, the distance to these clouds would place them near the edge of the Local Bubble, suggesting that they may be part of that structure rather than the main component of the Serpens clouds, which is far more distant \citep{ONeill24}.

\subsubsection{Serpens Northeast and South/W40}

Serpens Northeast contains an extensive network of filamentary structures oriented largely parallel to the galactic plane. This region is relatively unstudied, and as such, most of its HH objects are new to this publication. Serpens South, which contains the active star-forming region and highly-extinguished H II region W40, is located to the galactic west, and appears nearly contiguous with Serpens Northeast in dust maps like \citet{Bayestar19}. In total, we identify 21 HH objects, which are associated with 9 candidate progenitors. We list these below:

\noindent \textbf{HH108--109:} \citet{1992A&A...256L...1R,2000A&A...364..625S} identify \object{IRAS 18331-0035} as the driving source for this outflow.

\noindent \textbf{HH1284, HH1286, HH1289, HH1290, HH1291, HH1295 \& HH1297}:  All of these sources lie along the northwest edge of the \object{W40} complex.  They are roughly colinear and could possibly originate from the same progenitor. 
If so, the YSOs \object{IRAS 18264-0143} and \object{IRAS 18268-0140}, both of which have protostellar disks \citep{2024ApJ...973..138H}, are plausible candidates.  The orientation of the disk for IRAS~18264-0143, implying a stellar rotation axis aligned with the direction of the outflow, argues in its favor, in which case HH1284 and HH1297 would be the termini of the outflow.  Both are about 30\arcmin\ from IRAS~18264-0143, which is $\sim$4~pc at this distance.

\noindent \textbf{HH1306--1310:}  These objects lie along the perimeter of the dark nebula \object{Dobashi 1306}.  Embedded within this nebula is the red IR source \object{IRAS 18358-0112}, as well as several YSO candidates \citep[e.g.,][]{2021ApJS..254...33K}.  The positions of these sources indicate multiple progenitors.

\noindent \textbf{HH1300 \& HH1302:}  These objects lie along the northern edge of the dark nebula \object{Dobashi 1224}.  Close to these sources, and embedded in the nebula, is the YSO \object{HGBS J183629.6-022159} \citep{2015A&A...584A..91K}, about 5\arcmin\ to the southwest. 

\noindent \textbf{HH1305:}  Near this object there are three red YSO candidates, embedded in the dark nebula \object{Dobashi 1287} about 4.5\arcmin\ to the northwest: \object{SPICY 87918}, \object{SPICY 87939} and \object{SPICY 87984} \citep{2021ApJS..254...33K}.

\noindent \textbf{HH1312:}  This source is seen superimposed on the {Dobashi 1403} cloud and is only 18\arcsec\ from the YSO \object{IRAS 18383+0059}, which is the assumed source.

\noindent \textbf{HH1313:}  This object is not close to any dark nebula, nor any known YSOs.  The nearest YSO is IRAS~18383+0059, the likely progenitor for HH1312.

\noindent \textbf{HH1314:}  The dark nebula \object{Dobashi 1411} is to the southwest of this object. There are no known YSOs or red IR sources embedded within this nebula, or within 30\arcmin\ of this source.

\noindent \textbf{HH1315:}  This object lies to the northwest of the dark nebula \object{Dobashi 1410}; however, there are no known YSOs or red IR sources embedded within this nebula, or within 30\arcmin\ of this source.


All sources in Serpens Northeast have \citet{Edenhofer24} dust curves with peaks around 480~pc, without any secondary peaks that could cast doubt on this distance. This includes the potential progenitor HGBS J183629.6-022159, which is beyond the extent typically attributed to this cloud at $(l,b) = (29.1197, 2.2415)$, but has a dust peak at 479~pc. W40's proposed progenitors show a dust peak at 469~pc, generally in agreement with a distance of 460~pc from \citet{Herczeg19}. Using astrometry of water maser sources with the Very Long Baseline Array, \citet{2023A&A...673L...1O} finds a $440.7 \pm 4.6$ distance to Serpens South, which is often grouped with W40 \citep{Herczeg19}. However, \citet{Winston18} placed Serpens South in the foreground of W40, which could account for the discrepancy between the maser astrometry and dust peaks. One candidate progenitor in Ser NE has a \textit{Gaia} distance of $585^{+88}_{-65}$~pc, although the source that this is attributed to is one of the three candidate progenitors to HH1305. The measurement is therefore uncertain while also lacking a clear connection with the HH object, so we favor the distance from the dust maps. 

\subsubsection{Serpens Main}

The Serpens Main cloud is a well-studied star-forming region that is home to several known HH objects. In this work, we discover 8 new HH objects -- along with 12 previously known ones -- distributed across several components of the cloud. These objects are as follows: 

\noindent \textbf{HH106-107:}  These objects are co-linear with \object{IRAS 18269+0116}, a YSO long suspected to be the source \citep{1993A&A...273..221R}.  We identify a new faint component of HH107, $\sim$50\arcsec\ to the north of the bright lobe, and label it HH107B.

\noindent \textbf{HH455-460:}  These objects were co-discovered by \citet{1999MNRAS.309..141D} and \citet{1999A&A...347..565Z}.  The morphology of HH455 indicates a jet oriented at PA=-33\arcdeg, which points in the direction of Ser-emb 24 \& 29 \citep{2009ApJ...692..973E}. We do not detect \stwo\ line emission at the locations of HH456 \& 457 (Fig.~\ref{fig:hh_cutouts}).  The position of HH456 corresponds to a star, and no emission of any kind is detected at the position of HH457, suggesting that HH objects are not present at these locations. \citet{1999A&A...347..565Z} posit that the morphologies and positions of HH458 and HH460 indicate they are part of the same outflow.  They are aligned with the Serpens triple radio source and IRAS~18273+0113, suggesting this is the progenitor.  However, the morphologies and locations of HH458-460 do not indicate the source(s), as multiple YSO candidates lie within the cloud.

\noindent \textbf{HH476-478:} This source, discovered by \citet{1999A&A...347..565Z}, was identified by \citet{2016A&A...587A..75D} as part of a molecular hydrogen emission-line object (MHO) that they label as \object{MHO 2235}. HH476 is likely associated with the class I protostar \object{Ser-emb 11}, which lies 2.2\arcmin\ to the northeast  \citep{2009ApJ...692..973E}.  Alternatively, \object{MMS1} may be a potential progenitor \citep{2006A&A...458..789D}. HH477 appears to be aligned with the HH106-107 outflow, and may be associated with it.  \object{IRAS 18269+0116} is the closest YSO to HH477, but the morphology of this HH object suggests that its progenitor is more likely to be located in the main cluster. HH478 is clearly embedded in the Serpens reflection nebula and its driving source is expected to be located therein.

\noindent \textbf{HH1285 \& HH 1287:}  These are about 21\arcmin\ and 18\arcmin\ to the south of Ser-Emb~11, the protostar thought to be driving the MHO~2235 outflow that includes HH476.  This protostar may also be driving these sources, as they are the nearest candidates.  However, they are $\sim$10 times further away from Ser-Emb~11 than HH476. Alternatively, the YSO candidate \object{HBC 667} is closer and might be the progenitor for one of these sources.

\noindent \textbf{HH1288:}  HH1288 appears to be associated with the same outflow that contains \object{HH 106}, \object{HH 107}, and possibly \object{HH 477}.  

\noindent \textbf{HH1292--1294 \& HH1296:}  These faint HH objects are embedded in or near the Serpens cloud core, obfuscating their progenitors.  Their proximity to HH458, HH459, and HH478 suggests common origins, although this is unclear.  There are several notable YSOs with outflows in this region \citep[e.g.,][]{2024ApJ...973..138H}.  

\noindent \textbf{HH1298:}  This source is about 50\arcmin\ southeast of the Serpens cloud core, and may be associated with one of the outflows from this region.  Alternatively, it might be associated with \object{IRAS 18301+0035}, which lies 4.6\arcmin\ to the south. 


Aside from one HH object in an overdensity to the galactic south relative to the rest of the cloud, all progenitors and candidate progenitors have \citet{Edenhofer24} dust peaks at $d\sim$440~pc. This agrees closely with two sources for which we have \textit{Gaia} distances \citep{2021AJ....161..147B}: \object{IRAS 18269+0116} at $442.2 \pm  28.6$~pc, and \object{2MASS J18300341+0116191} at $443.1 \pm 11.5$~pc. The southern object appears to be more distant, with a peak at $d\sim$487~pc. This would place it at a distance closer to Serpens Northeast, despite its projected position closer to the other Serpens Main clouds in the plane of the sky. However, the \citet{Edenhofer24} maps show a bimodality in the dust distribution between clouds at $\sim$440~pc and $\sim$490~pc, so an origin in the nearer clouds connected to Serpens Main remains possible. Aside from this outlier, our distance to sources in Serpens Main is in close agreement with the 445~pc distance cited in \citet{Herczeg19}.

\subsubsection{Scattered Sources near Serpens}

Eight of the HH objects in the western part of our survey near Serpens do not have strong ties to any major molecular clouds. We list these objects below:

\noindent \textbf{HH1279--1281}:  These three HH objects are embedded in the dark nebula Dobashi 1139 \citep{Dobashi11}.  They are linearly aligned with \object{IRAS 18197-0158}, strongly suggesting it is the progenitor for all three.

\noindent \textbf{HH1299, HH1301, HH1303 \& HH1304:}  These objects are embedded in a large HII emission nebula that contains over 100 YSOs.  HH1301 has a giant bow-shock structure similar to \object{HH401}; its morphology, and the shape of HH1304, suggest the origin is to the west.  HH1299 and HH1303 are also likely part of this outflow, { as is MHO~2282 \citep{2012MNRAS.421.3257I}}.  There are multiple YSOs in the nearby NIR bubble \object{N37} \citep{2006ApJ...649..759C} that are plausible progenitors \citep{2016ApJ...833...85B}.  \object{SSTGLMC G025.2255+00.2884}, and further to the west \object {IRAS 18327-0645}, are well aligned and represent the most likely progenitor candidates.  

\noindent \textbf{HH1311:}  This compact object lies along the northern edge of the dark nebula \object{SDC G25.005-0.635}.  Unfortunately its morphology doesn't suggest a direction to the progenitor. \object{2MASS J18391236-0720158} is the nearest YSO candidate.  { Interestingly, HH1311 is coincident with IFO~99 \citep{2024MNRAS.528.4657K} and MHO~2291 \citep{2012MNRAS.421.3257I}.}

The HH objects HH1279--1281 and their probable progenitor \object{IRAS 18197-0158} are located to the galactic north of Serpens, near the intersection of the near and far components. While dust maps indicate the presence of both a near and a far dust component along this line of sight, the \citet{Bayestar19} dust maps show that the near component in this region is generally broad, while the far component has a distribution consistent with the clouds that this source is embedded in. We therefore base our distance to these sources on this far extinction peak. 

The remaining two sources are located to the galactic south of Serpens-Near, within the Scutum inset map in Fig.~\ref{fig:otherinsets}. \citet{Edenhofer24} shows three dust peaks in the direction of Scutum, most notably one at a distance connected with the main component of the Serpens Molecular Cloud, and one consistent with Serpens Near and Serpens Cirrus \citep{Herczeg19}. The \citet{Bayestar19} maps suggest that the dust in the region is concentrated in the distance range  $200 \lesssim d \lesssim 300$~pc, and place nearly all major dust concentrations in the area within this range. We therefore conclude that this region is most likely connected to Serpens Near, although given the ambiguity in the \citet{Edenhofer24} dust maps, we cannot rule out an origin in a different cloud along the same sight line.



\subsection{LDN~673 / SCYA-59}

\begin{figure}[ht]
\plotone{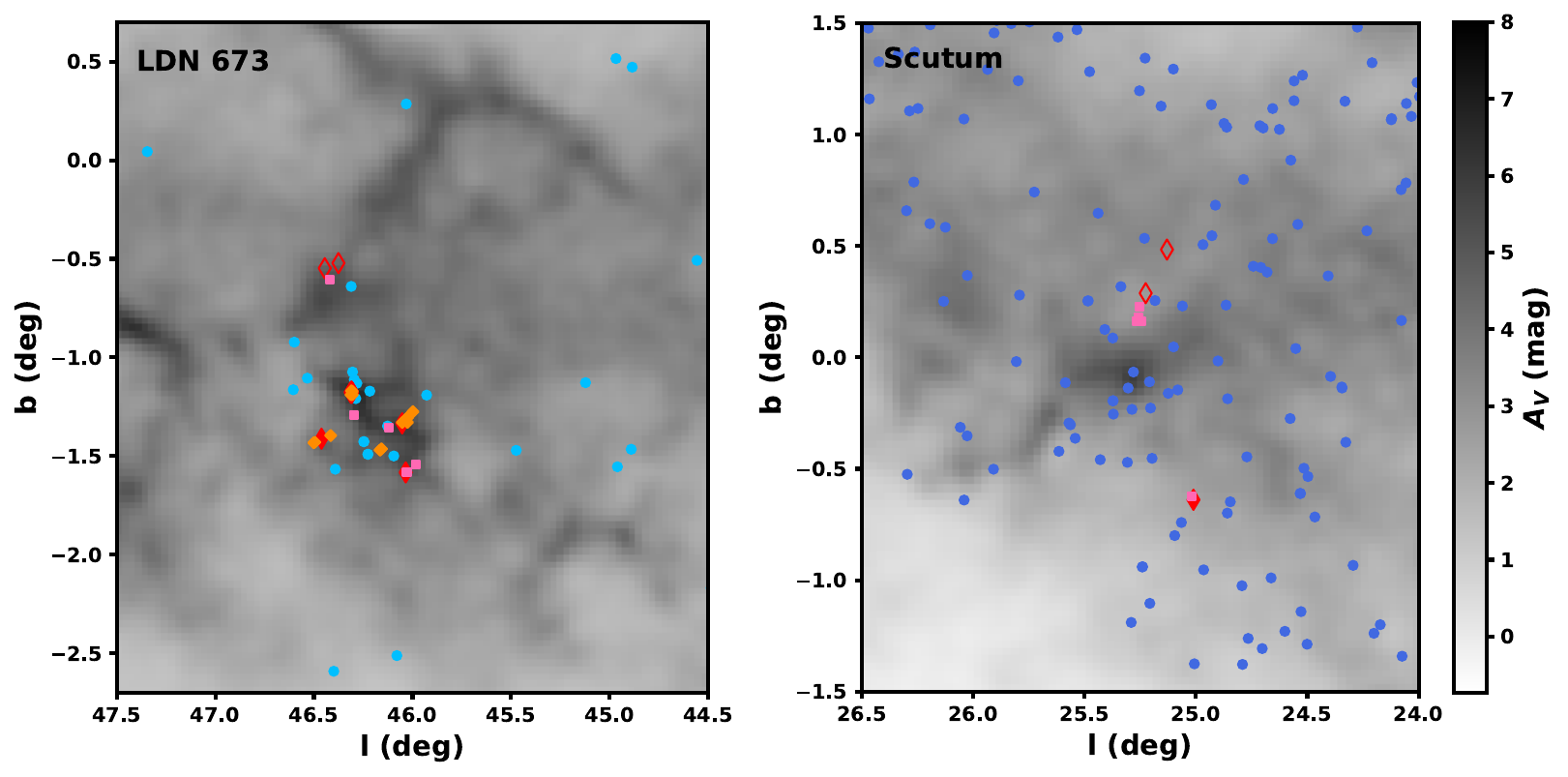}
\caption{Same as Fig.~\ref{fig:aquilarift}, but for LDN~673 (left), and for Scutum (right), showing spatially disconnected HH objects to the galactic south of the Serpens Complex.
\label{fig:otherinsets}}
\end{figure}

LDN~673, located to the far east of the Aquila rift, is a compact and isolated cloud in which 15 HH objects were already known; and here we report 5 new discoveries. The individual sources in LDN~673 are as follows:

\noindent \textbf{HH32 \& HH332:}  The progenitor for this well-studied outflow \citep[e.g.,][]{2020AJ....160..165H} is the T-Tauri multiple star system AS~353. As discussed in \citet{2018ApJ...852...13R}, there is a third IR-component in AS~353 that might be a driving source.  { As discussed by  \citet{2024MNRAS.528.4657K}, IFO~153 is coincident with HH32.}

\noindent \textbf{HH250:}  The progenitor for this outflow is thought to be \object{IRAS 19190+1048}, a binary system of two Class I YSOs \citep{2019A&A...628A..97C}.  { IFO~155 is coincident with this source \citep{2024MNRAS.528.4657K}.}

\noindent \textbf{HH1183-1194:}  \citet{2018ApJ...852...13R} first reported the discovery of these objects.  They also identified eight IR sources that could be the progenitor(s) for HH1183-1186.  HH1187-1191 are part of the same outflow as HH32 and HH332, and HH1192-1194 are likely driven by the YSO SSTc2d~J192134.8+112123 \citep{2008ApJS..179..249D}.

\noindent \textbf{HH1357:}  The morphology of this source suggests the progenitor is to the northwest.  The source \object{IRAS 19160+1139} lies 3.4\arcmin\ away in that direction but little is known about it.  Alternatively, the YSO candidate \object{IRAS 19158+1136} is 5.2\arcmin\ directly to the west and should be considered a potential progenitor.  

\noindent \textbf{HH1358:}  This extended HH object is quite faint because it is embedded in \object{LDN~673}.  It appears to be an extension of the northeastern flow from \object{HH 32}, which originates in AS~353.

\noindent \textbf{HH1359:}  This object is deeply embedded in LDN~673 and is barely visible.  It lies about 6.5\arcmin\ to the southwest of the HH1183-86 group, for which potential progenitors are discussed by \citet{2018ApJ...852...13R}. 

\noindent \textbf{HH1360--1361:}  These appear to be extensions of the HH250 outflow.


We show the locations of HH objects and their progenitors in the left panel of Fig.~\ref{fig:otherinsets}. All progenitors clearly align with local dust overdensities within this wispy, filamentary cloud. LDN~673 has been recently shown to house a small young stellar association, identified as group 369 in \citet{Prisinzano22} and as SCYA-59 in \citet{Kerr23}, where it is given an essentially newborn age of $\sim$1~Myr. AS~353, the proposed progenitor for HH32 and HH1358, has a high-quality \textit{Gaia} entry that is grouped into SCYA-59 in \citet{Kerr23}. The extinction peaks from the \citet{Edenhofer24} dust curves align closely with the SCYA-59 members for all progenitors, as shown in Fig.~\ref{fig:distprofs}. We summarize the distance to the candidate progenitors of LDN~673 HH objects in Table \ref{tbl:progens}, alongside those of other Serpens-Aquila regions.

\startlongtable
\begin{deluxetable*}{lllllllll}
\tablecolumns{7}
\tablewidth{0pt}
\tablecaption{\label{tbl:progens} Progenitor Candidates and Their Distances}
\tablehead{\colhead{Stars} & \colhead{RA (2000)} & \colhead{Dec} & \colhead{$l$} & \colhead{$b$} & \colhead{d (pc)} & \colhead{$\sigma_d^-$} & \colhead{$\sigma_d^+$} & \colhead{d src\tablenotemark{a}}}
\startdata
IRAS 18148-0440 & 274.3748 & -4.6609 &      24.8944 &      5.35275 & 241  & 25 
 & 13  & dust \\
IRAS 18197-0158 & 275.5929 & -1.9498 &      27.8314 &      5.58965 & 475  & 18 
 & 40  & dust \\
2MASS J18251133-0258532 & 276.2972 & -2.9815 &      27.8617 &      5.56593 &
 255  & 9  & 13  & dust \\
2MASS J18251332-0259549 & 276.3055 & -2.9986 &      27.8689 &      5.55934 &
 255  & 9  & 13  & dust \\
IRAS 18264-0143 & 277.272 & -1.6991 &      25.2522 &      4.01779 & 469  & 25  &
 14  & dust \\
\enddata
\tablecomments{Only the first five entries of this table are shown here.  The progenitor candidates and their distances are published in their entirety in machine-readable format.}
\tablenotetext{a}{Distances determined either from Gaia, dust maps \citet{Edenhofer24}, or maser-based parallax 
\citep{Zhang09}.} 
\end{deluxetable*}

\section{Candidate Progenitor Classifications} \label{sec:progenitors}

In order to better assess the nature of the candidate progenitors, we used {\it Wide Field Infrared Explorer (WISE)}  broad-band colors to estimate YSO class.  The {\it WISE} mission mapped the full sky in 4 mid-IR bands:  3.4 (W1), 4.6 (W2), 12 (W3), and 22 \micron\ (W4) \citep{wright2010}.  Though the original mission objective was to identify and measure the most luminous galaxies in the Universe, the {\it WISE} images and catalogs have become indispensable to other astrophysical studies, including star formation and the evolution of YSOs.   

\begin{figure*}[ht!]
\plotone{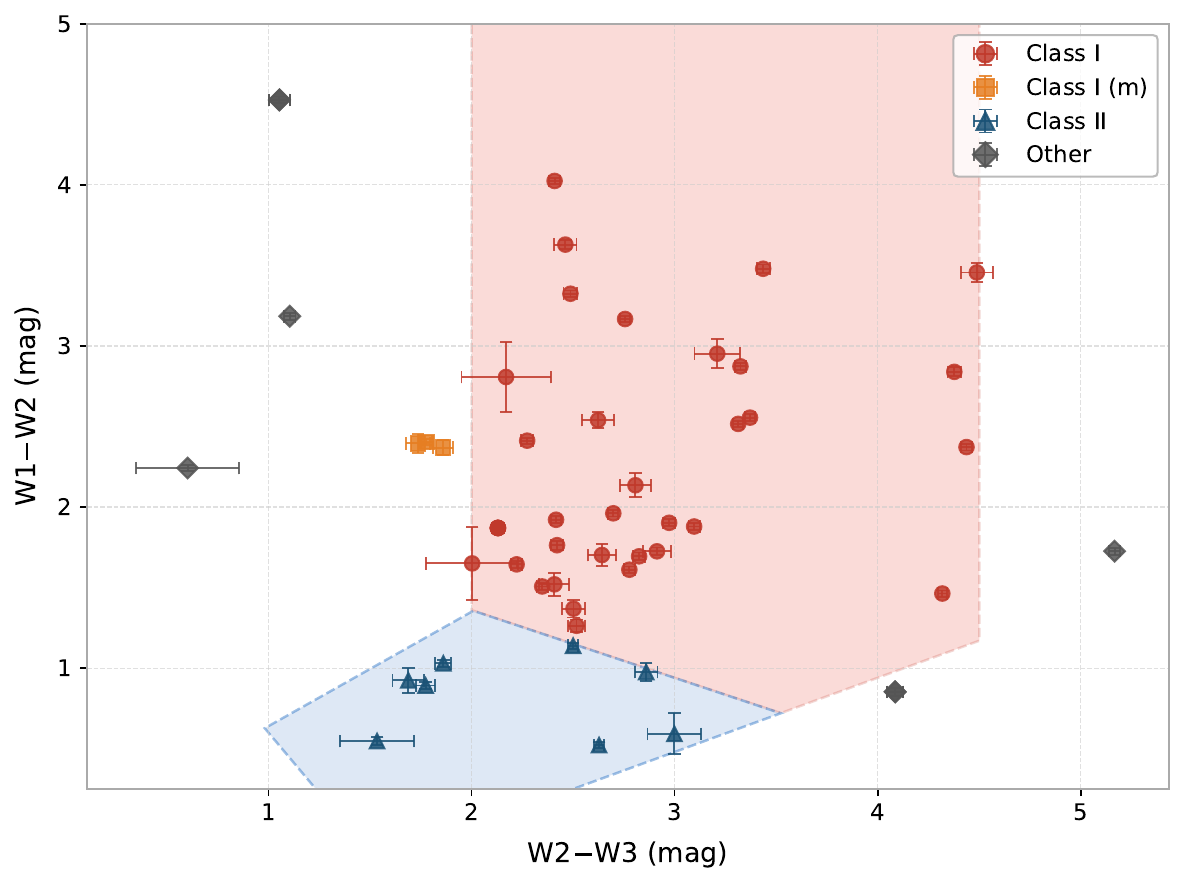}
\caption{{ {\it WISE} color-color diagram for the candidate progenitors in Table~\ref{tbl:progenitors_class}. Shaded areas indicate the YSO class identification scheme defined in \citet{fischer2016}.  Class I ``marginal'' sources are depicted in orange -- see text for discussion.}
\label{fig:wise_cc_ysos}}
\end{figure*}

Using the  coordinates from Table~\ref{tbl:progens} for each candidate, we queried the AllWISE catalog \citep{cutri2013} at IRSA\footnote{\href{https://irsa.ipac.caltech.edu/data/WISE/docs/release/AllWISE/index.html}{https://irsa.ipac.caltech.edu/data/WISE/docs/release/AllWISE/index.html}} for matches using a cone search with radius $10$\arcsec\ and one-to-one matching.  Our search returned  58 of the 66 candidate progenitors.  We then calculated the ($W1 - W2$) and ($W2 - W3$) colors for  each source, and applied the color-color YSO classification scheme used by \citet{fischer2016} in Canis Major, which is a slight modification to the original scheme first described by \citet{koenig2014}.  This scheme is based on {\it WISE} colors of known YSOs in Taurus, extragalatic sources, and other contaminants, and hence is somewhat more reliable than classification based on spectral slope in the IR \citep{lada1987} or bolometric temperature \citep{chen1995}.  Four of the candidate progenitors are missing photometry for at least one of the bands and hence could not be classified.  The matching sources, {\it WISE} colors, and YSO class determinations are shown in Table~\ref{tbl:progenitors_class}, and on a color-color plot in Fig.~\ref{fig:wise_cc_ysos}.  Three sources are just below the \citet{fischer2016} $(W2 - W3) = 2$ cutoff for Class~I sources. Since it seems likely that these are in fact YSOs, we have marked these as ``marginal'' Class~I [``I (m)''].  Of the 58 sources found in AllWISE,  39 appear to be Class~I (including the 3 marginal cases), 8 are Class~II, 7 have colors that are inconsistent with YSOs (``other''; but see individual source discussions below), and 4 have insufficient photometry for classification.  We have not tried to correct the colors for extinction since this effect is relatively small in the mid-IR, nor would it affect the overall YSO classification much.  It is possible, however, that some of the Class~I sources with low ($W1 - W2$) values could in fact be heavily extincted Class~II sources.  

\startlongtable
\begin{deluxetable*}{lrrlcll}
\tablecaption{Progenitor Candidates:  WISE colors and YSO classification\label{tbl:progenitors_class}}
\tablewidth{0pt}
\tablehead{
\colhead{Source} &
\colhead{$[W1]-[W2]$} &
\colhead{$[W2]-[W3]$} &
\colhead{Class } &
\colhead{$\alpha$} &
\colhead{C2D Type} &
\colhead{Class}\\
\colhead{} &
\colhead{(mag)} &
\colhead{(mag)} &
\colhead{(WISE)} &
\colhead{} &
\colhead{} &
\colhead{(C2D)}
}
\startdata
IRAS 18148-0440 & 3.184 & 1.105 & other & 1.30 & YSOc\_red & 0/I/F \\
IRAS 18197-0158 & 1.702 & 2.642 & I & \nodata & \nodata & \nodata \\
2MASS J18251133-0258532 & 2.839 & 4.377 & I & \nodata & \nodata & \nodata \\
2MASS J18251332-0259549 & 1.644 & 2.222 & I & \nodata & \nodata & \nodata \\
IRAS 18264-0143 & 2.403 & 1.778 & I (m) & \nodata & \nodata & \nodata \\
IRAS 18268-0140 & 2.516 & 3.313 & I & \nodata & \nodata & \nodata \\
Ser-emb 7 & 3.167 & 2.756 & I & 1.36 & YSOc\_red & 0/I/F \\
MMS 1 & 4.527 & 1.054 & other & 2.56 & red & 0/I \\
IRAS 18269+0116 & 1.520 & 2.408 & I & 0.26 & YSOc\_star+dust(IR1) & F/II \\
Serpens Cloud Core & \nodata & \nodata & \nodata & -0.26 & red & 0/I \\
IRAS 18301+0035 & 0.892 & 1.774 & II & \nodata & \nodata & \nodata \\
IRAS 18327-0645 & 3.456 & 4.488 & I & \nodata & \nodata & \nodata \\
SSTGLMC G025.2255+00.2884 & 1.030 & 1.861 & II & \nodata & \nodata & \nodata \\
HGBS J183629.6-022159 & 2.555 & 3.371 & I & \nodata & \nodata & \nodata \\
SPICY 87984 & \nodata & \nodata & \nodata & \nodata & \nodata & \nodata \\
SPICY 87939 & 1.725 & 2.913 & I & \nodata & \nodata & \nodata \\
SPICY 87918 & 2.539 & 2.623 & I & \nodata & \nodata & \nodata \\
IRAS 18358-0112 & 4.024 & 2.409 & I & \nodata & \nodata & \nodata \\
2MASS J18391236-0720158 & 0.547 & 1.535 & II & \nodata & \nodata & \nodata \\
IRAS 18383+0059 & 1.763 & 2.421 & I & \nodata & \nodata & \nodata \\
IRAS 18440+0023 & 1.961 & 2.698 & I & \nodata & \nodata & \nodata \\
IRAS 18454+0250 & 0.975 & 2.860 & II & \nodata & \nodata & \nodata \\
SPICY 92856 & 1.467 & \nodata & \nodata & \nodata & \nodata & \nodata \\
2MASS J18494869+0006072 & 0.314 & \nodata & \nodata & \nodata & \nodata & \nodata \\
SPICY 92905 & 3.629 & 2.462 & I & \nodata & \nodata & \nodata \\
SPICY 92890 & 0.118 & \nodata & \nodata & \nodata & \nodata & \nodata \\
IRAS 18484+0237 & 1.611 & 2.777 & I & \nodata & \nodata & \nodata \\
GGD30IR & 2.395 & 1.736 & I (m) & \nodata & \nodata & \nodata \\
MSX6C G034.0126-00.2832 & 0.853 & 4.086 & other & \nodata & \nodata & \nodata \\
MSX6C G034.0500-00.2977 & 0.853 & 4.086 & other & \nodata & \nodata & \nodata \\
HIGALBM G036.9422+01.1219 & \nodata & \nodata & \nodata & \nodata & \nodata & \nodata \\
IRAS 18528+0400 & 1.263 & 2.517 & I & \nodata & \nodata & \nodata \\
HIGALBM G034.3721-00.5424 & 1.902 & 2.973 & I & \nodata & \nodata & \nodata \\
AGAL G034.196-00.592 & 1.463 & 4.318 & I & \nodata & \nodata & \nodata \\
HIGALBM G034.3395-00.6940 & 1.507 & 2.348 & I & \nodata & \nodata & \nodata \\
IRAS 18555+0056 & 2.952 & 3.210 & I & \nodata & \nodata & \nodata \\
IRAS 18550+0104 & 1.650 & 2.003 & I & \nodata & \nodata & \nodata \\
IRAS 18542+0114 & 2.807 & 2.170 & I & \nodata & \nodata & \nodata \\
BCM2016 8a/G35.2-0.74N & 3.479 & 3.436 & I & \nodata & \nodata & \nodata \\
IRAS 18558+0047 & 1.139 & 2.500 & II & \nodata & \nodata & \nodata \\
IRAS 18566+0048 & 2.873 & 3.324 & I & \nodata & \nodata & \nodata \\
HIGALBM G038.6940+00.4527 & 1.921 & 2.416 & I & \nodata & \nodata & \nodata \\
MSX6C G038.8750+00.3088 & \nodata & \nodata & \nodata & \nodata & \nodata & \nodata \\
SPICY 96721 & 0.592 & 2.999 & II & \nodata & \nodata & \nodata \\
IRAS 18589+0511 & 1.879 & 3.096 & I & \nodata & \nodata & \nodata \\
CORNISH G039.1956+00.2255 & 1.726 & 5.166 & other & \nodata & \nodata & \nodata \\
HBC 684 & 1.369 & 2.502 & I & \nodata & \nodata & \nodata \\
IRAS 19045+0518 & 2.135 & 2.807 & I & \nodata & \nodata & \nodata \\
IRAS 19160+1139 & 2.368 & 1.860 & I (m) & \nodata & \nodata & \nodata \\
IRAS 19158+1136 & 2.372 & 4.437 & I & \nodata & \nodata & \nodata \\
AS 353 & 0.924 & 1.689 & II & 0.11 & star+dust(IR2) & II \\
IRAS 19190+1048 & 1.694 & 2.825 & I & \nodata & \nodata & \nodata \\
IRAS 18331-0035 & 2.193 & \nodata & \nodata & \nodata & \nodata & \nodata \\
Ser-emb 8 & 3.324 & 2.487 & I & 1.37 & YSOc\_red & 0/I/F \\
IRAS 18273+0113 & 4.527 & 1.054 & other & 2.65 & YSOc\_red & 0/I/F \\
Ser-emb 17 & \nodata & \nodata & \nodata & 1.70 & YSOc\_red & 0/I/F \\
2MASS J18300341+0116191 & 0.522 & 2.628 & II & -0.57 & YSOc\_star+dust(IR3) & II \\
HH1183–1186 IRS1 & 1.870 & 2.130 & I & 1.34 & star+dust(MP1) & II \\
HH1183–1186 IRS2 & 1.870 & 2.130 & I & 0.98 & YSOc\_red & 0/I/F \\
HH1183–1186 IRS3 & 2.242 & 0.602 & other & -0.68 & YSOc & I/F/II \\
HH1183–1186 IRS4 & 1.870 & 2.130 & I & 0.16 & YSOc\_star+dust(IR2) & F/II \\
HH1183–1186 IRS5 & 1.870 & 2.130 & I & -0.76 & YSOc\_red & 0/I/F \\
HH1183–1186 IRS6 & \nodata & \nodata & \nodata & -0.39 & YSOc & I/F/II \\
HH1183–1186 IRS7 & 2.412 & 2.274 & I & 0.45 & YSOc & I/F/II \\
HH1183–1186 IRS8 & \nodata & \nodata & \nodata & -0.86 & star+dust(IR4) & II \\
SSTc2d J192134.8+112123 & \nodata & \nodata & \nodata & 0.72 & red & 0/I \\
\enddata
\tablecomments{Columns: Source name; WISE color indices $[W1]-[W2]$ and $[W2]-[W3]$ (mag); 
WISE classification; spectral index $\alpha$; object type and classification from the c2d survey.}
\end{deluxetable*}

We also searched the Cores2Disks catalogs \citep{evans2009} because a couple of the sub-regions in Serpens-Aquila were also included in those surveys. The C2D project identified YSOs in a number of well-studied star forming regions using Spitzer IRAC and MIPS photometry.  The C2D classifications are likely to be more reliable than those derived from WISE colors given the additional, longer wavelength MIPS band data and the higher sensitivity of Spitzer. The query returned 19 unique matches from the CLOUDS and CORES catalogs, all of which are YSOs.  We also estimated traditional YSO class for each source using Figs.~8 \& 9 of \citet{evans2009}, which are based on bolometric temperature ($T_{bol}$) \citep{chen1995} and spectral slope in the IR ($\alpha_{IR}$) \citep{lada1987}.  The C2D object type, $\alpha_{IR}$, and inferred YSO Class are also shown in Table~\ref{tbl:progenitors_class}.  The object type defintions are:
\begin{itemize}
\item{\tt Star} -- photometry is consistent with stellar photosphere.
\item{\tt YSOc} -- photometry is consistent with YSO SED (sometimes coupled with a {\tt star+dust} or {\tt red} identification, see below). 
\item{\tt red}  -- excess flux at 24\micron.
\item{\tt PAH\_em}  -- excess flux at 8\micron
\item{\tt star+dust(BAND)} --  stellar photosphere + excess flux at given band.
\end{itemize}
All 19 C2D matches have Spitzer colors consistent with YSOs; but as can be seen from the table, the specific YSO class cannot be determined except for some of the {\tt star+dust} cases.  Of the 14 overlapping sources with classifications for both WISE and C2D, only 7 are consistent with each other.   Three of the ``other'' WISE sources are candidate YSOs in the C2D catalog, which is not unexpected given the additional longer wavelength MIPS bands used in the C2D survey.  For the remaining sources that disagree, the WISE colors tend to infer Class~I while C2D infers Class~II.  If the C2D really is more reliable, then we might expect some of the other Class~I sources identified with the AllWISE data to actually be Class~II as well.  

In summary, WISE color-color analysis as well as C2D identifications indicate that 55 of the 66 candidate progenitors are likely to be YSOs.  The remaining 11 sources are: 
\begin{itemize}
\item SPICY 87984: Source present in AllWISE images, but listed in the AllWISE Reject Catalog.  Identified as a candidate YSO in the SPICY survey \citep{kuhn2021}.
\item MSX6C G038.8750+00.3088: Source present in AllWISE images, but listed in the AllWISE Reject Catalog.  Identified as candidate YSO by \citet{urquhart2009}.
\item HIGALBM G036.9422+01.1219: mm-source with a very faint detection in WISE Band 4 \citep{elia2017}; likely a Class~0 source.
\item MSX6C G034.0126-00.2832 \& MSX6C G034.0500-00.2977:  Classified as ``other'' using WISE colors; identified as candidate massive YSOs by \citet{munar-adrover2011}.
\item CORNISH G039.1956+00.2255:  Classified as ``other'' from WISE colors; identified as a YSO by \citet{2009A&A...501..539U}.
\item SPICY~92856 \& 92890:  insufficient WISE photometry for classification; faint YSO candidates from the SPICY survey \citep{kuhn2021}.
\item 2MASS J18494869+0006072: insufficient WISE photometry for classification; identified as a YSO candidate (SPICY 92854) by \citet{kuhn2021}.
\item IRAS 18331-0035:  insufficient WISE photometry for classification; SED\footnote{SIMBAD Vizier Photometry} peaks in the far-IR suggesting this is a Class~0 source; \citet{1992A&A...256L...1R} identify this object as the driving source for HH108 \& 109.
\end{itemize}
In summary, nearly all of the candidate progenitors are likely to be YSOs based on the available {\it WISE} photometry and/or previous studies.

\section{Overview of Star-Forming Regions Traced by HH Objects} \label{sec:disc}

\begin{figure}[ht]
\plotone{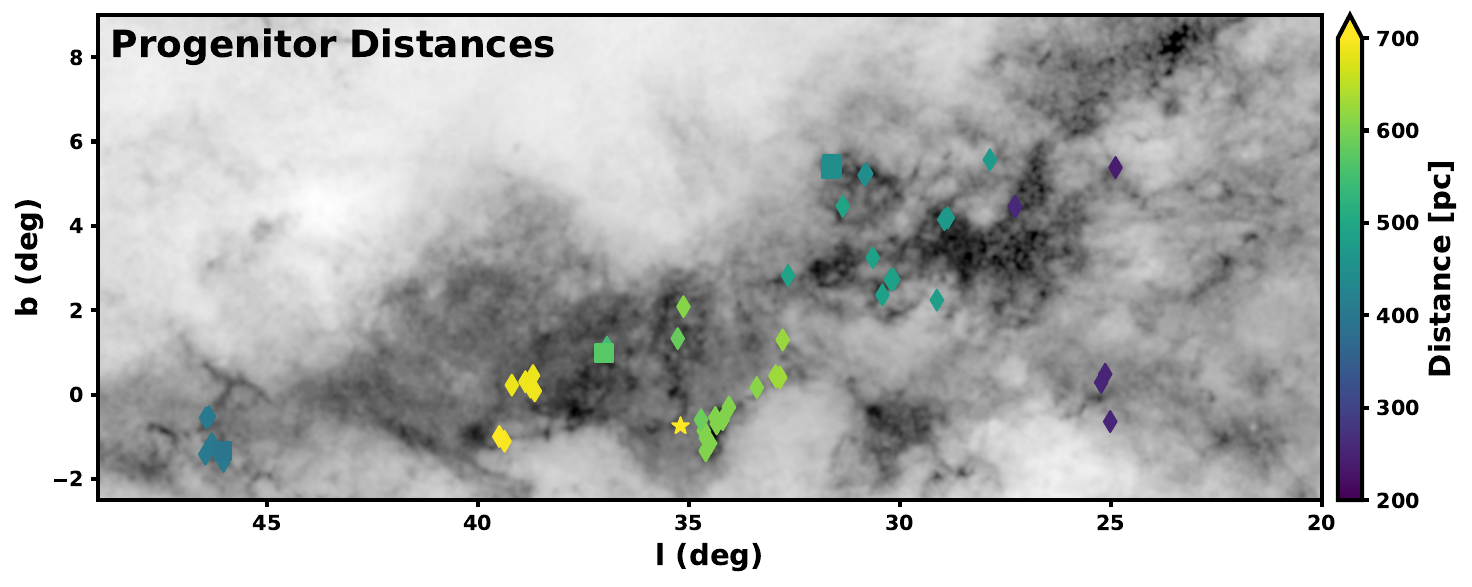}
\caption{Map of candidate progenitors to HH objects, colored by their distances, measured from their most probable dust peak (diamonds), Gaia distance (squares), or maser parallax (star icon) \citep{Zhang09}. The background is the same \citet{2005PASJ...57S...1D} dust map shown in Fig.~\ref{fig:hh_loc}. 
\label{fig:distmap}}
\end{figure}

In Fig.~\ref{fig:distmap} we provide an overview of the distances to the candidate progenitors in the Serpens-Aquila Rift region. We show the locations of these HH objects in the plane of the galaxy in Fig.~\ref{fig:dust3d}, with a full 3D version available in the online-only figure. These maps reveal at least five spatially distinct regions in which HH objects verify the presence of active star formation: (1) the main Serpens Complex, which overlaps with the stellar population from \citet{Kerr23}; (2) a near component that corresponds to an extinction wall at $\sim$250~pc \citep{Straizys03}; (3) LDN~673/ SCYA-59; and (4-5) two more distant components in Aquila, with one consisting primarily of LDN~617 and the other corresponding to clouds around 700~pc away. HH1337 is a notable outlier, which is associated with the G35.2-0.74N cloud at $\sim$2.2~kpc. Here we discuss each of these components, and what the gas, stars, and HH objects in the region tell us about their past, present, and future star formation. 

\subsection{Serpens Molecular Cloud}

The Serpens Molecular Cloud is the most widely-known population we identify here, containing Serpens Main, NE, and South/W40. It has therefore been the subject of numerous surveys of molecular tracers of dense gas and protostars, and is associated with large young stellar populations in both the \citet{Kerr23} and \citet{Prisinzano22} maps. We find that parts of Serpens Main are notably closer than the rest of the complex, with highly self-consistent distances of $443.2\pm0.6$~pc (mean and standard deviation) across both \textit{Gaia} and dust peak-based estimates. This is $\sim$40~pc closer than the rest of the complex, with consistent distances of 481$\pm$7~pc across all Serpens clouds outside the central component of Serpens Main. This difference is evident in Fig.~\ref{fig:dust3d}, where an overdensity in both extinction and stars extends from the main body of the Serpens cloud towards the HH objects in Serpens Main. Therefore, while these clouds are spatially distinct, both are fully consistent with the stars and gas that mark the past and probable future generations in Serpens. 

The stellar component in Serpens extends to far closer distances than the gas clouds, occupying much of an apparent cavity between the Serpens Molecular Cloud and Serpens Near. Toward the far edge of this cavity, the MWC 297 cluster (also known as Serpens Far-South; \citealt{Herczeg19}) forms a clear stellar overdensity. \citet{Herczeg19} marks this population as the oldest in the region, roughly following a 10 Myr PARSEC isochrone. This age and the population's position within the enclosing cavity suggest that the cavity is shaped by stellar feedback in this cluster, which drives its expansion and the compression of material in its shell. The location of the Serpens Molecular Cloud at the rear edge of this cavity suggests that feedback-driven compression initiated by the MWC 297 cluster may power current generations of star formation in Serpens Main \citep[e.g.,][]{Elmegreen77}. 

\subsection{Serpens Near}

Serpens Near is close enough to the sun that its position roughly aligns with the edge of the Local Bubble as proposed in \citet{Kerr26a}, so it may form a small part of that much larger structure. The nearest stellar population to Serpens Near that has been connected to the Local Bubble is Scutum North (ScuN; SCYA-70), which is offset just to the galactic west of the Scutum region at (l, b) = (23.6, 0.3). This region consists of three sub-components of distinct ages, with the two youngest components (collectively referred to as ScuN-Y), sharing velocities consistent with the expansion of the Local Bubble. The youngest component in ScuN-Y, ScuN-Cen has an age of 6 Myr, and therefore shows a compact stellar distribution consistent with minimal dispersal. ScuN-Cen is located within $\sim$20~pc of the heavily extinguished regions that contain Serpens Near's HH objects, making the cloud a plausible continuation of multi-generational star formation in Scutum North.

A connection with Serpens Near would require that its velocities are similar to ScuN-Cen, which has $v_r=-7.3$ km s$^{-1}$, or $v_{LSR} = 8.0$ km s$^{-1}$ \citep{Kerr26a}. CO gas in Serpens Near was mapped by \citet{Nakamura17} using a 1.85m telescope at the Nobeyama Radio Observatory, providing local standard of rest (LSR) velocities across the entire Serpens complex. The velocities related to Serpens Near and the more distant parts of the cloud did not cleanly separate in $v_{LSR}$ space, but both are consistent with ScuN-Cen, at 5 and 9 km s$^{-1}$. The higher-velocity component appears to correspond to Serpens Near, as the 9-12 km s$^{-1}$ window shows filamentary overdensities consistent with those that contain the region's HH objects, while the lower-velocity component is far more uniform in that same region, and shows far clearer overdensities in regions associated with the Serpens Molecular Cloud, especially W40 \citep{Nakamura17}. This supports Serpens Near's status as a continuation of star formation in ScuN, as its slightly higher radial velocities are expected for a later-forming stellar population in an accelerating bubble such as the Local Bubble. The location of Serpens Near, offset outward from the center of the local bubble relative to ScuN, supports this interpretation, as an accelerating cloud may be blown clear of its earlier stellar generations \citep{Posch23}.

\subsection{LDN~673}

\begin{figure}[ht]
\plotone{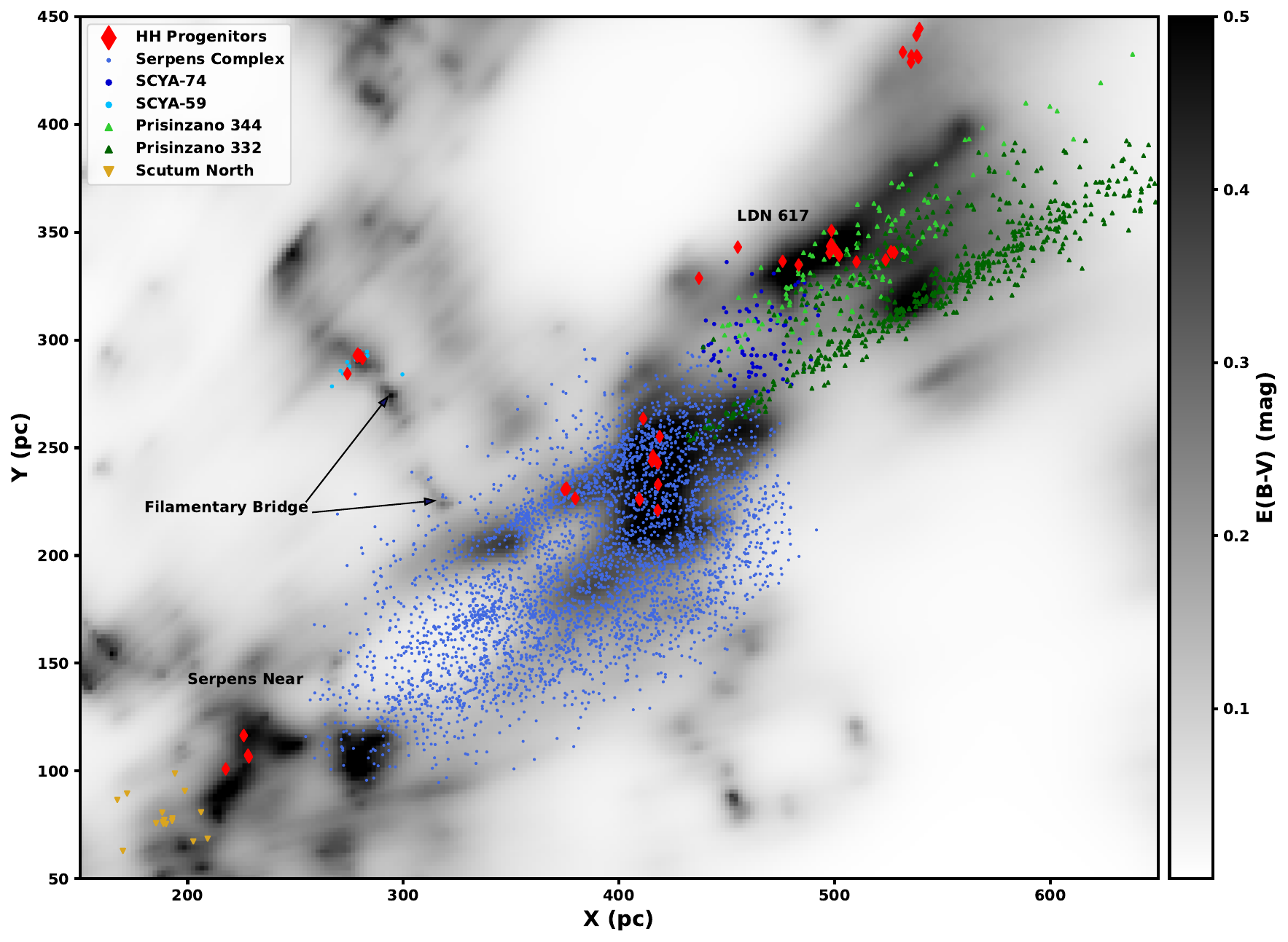}
\caption{Locations of Herbig-Haro object progenitors in XY galactic coordinates, overlaid on \citet{Edenhofer24} dust maps. We include Scutum North, SCYA-59, SCYA-74, and the Serpens Complex from \citet{Kerr23}, and group 332 and 344 from \citet{Prisinzano22} for reference. A 3D interactive version is available online.
\label{fig:dust3d}}
\end{figure}

On the opposite end of the complex, the much smaller but well-characterized LDN~673 region also hosts HH objects alongside young stellar populations containing several embedded sources. Its distance is similar to Serpens at $404.1 \pm 2.6$~pc, but its 15-20 degree separation, corresponding to nearly 200~pc, makes it unlikely that either population directly affects the other. However, the 3D \citet{Edenhofer24} dust maps available in Fig.~\ref{fig:dust3d} show an extended filamentary overdensity that stretches towards the main Serpens complex (labeled ``filamentary bridge'' in Fig.~\ref{fig:dust3d}). Investigating the coherence of this filament and its motions relative to Serpens may provide important insight into star formation in Serpens and the broader region. If this filament and Serpens have convergent motions, it may indicate an active cloud-cloud collision, which may impact future star formation in the Serpens Molecular Cloud.

\subsection{Aquila Rift}

The clouds that we group broadly into the Aquila Rift,  located between Serpens and LDN~673 on-sky, are highly inhomogeneous.  They contain two well-separated distance components, with the western component having $550 \lesssim d \lesssim 630$~pc, while the eastern objects have distances of nearly 700~pc. These components are separated by over 100~pc in 3D space, so we divide them into two broad cloud complexes that we hereafter refer to as the Western and Eastern Aquila Rift. The former complex contains clouds like LDN~617 and LDN~628, while the latter consists of clouds marked as AR-1 in Fig.~\ref{fig:aquilarift}. The distinction between these two clouds is strengthened by the $^{12}$CO gas velocities in \citet{2008hsf1.book...18P}, where most of the gas associated with the Eastern Rift has $v_{LSR} \simeq 15$~km~s$^{-1}$, while the Western Rift is closer to 10~km~s$^{-1}$. 

The western edge of the Western Aquila Rift contains components of SCYA-74 and \citet{Prisinzano22} group 344, with stellar overdensities that match with parts of LDN~617 and its adjoining clouds. Most of the HH objects in this region cling to the edge of a large low-extinction cavity visible in Fig.~\ref{fig:distmap}. At the center of this apparent cavity lies a sub-component of \citet{Prisinzano22} group 332, which also shares some overlap with SCYA-74. With a mean distance of 645~pc, this population has a distance within less than 20~pc of the most distant stars associated with LDN~617, so it may be the driver of a bubble that LDN~617 and its adjoining clouds form the edge of. A more exhaustive survey of the region including ages may better resolve the different stellar populations, and determine which populations might have dominated local stellar feedback and sculpted the dense clouds in the region.

Outside of LDN~617 and adjoining clusters, the rest of the Aquila Rift shows minimal connections to known stellar populations, in both the close and far components \citep{Prisinzano22, Kerr23}. This indicates that star formation may be only just beginning in these regions. However, we lack any additional information on the broader motions of stars in this region, and how they might relate to adjacent gas clouds. The upcoming work of the Roman Galactic Plane survey will cover most of this region, and it may therefore provide essential infrared coverage that will improve our understanding star formation here \citep{Paladini23}. 

\section{Conclusions} \label{sec:conclusions}

We have conducted a optical survey covering $\sim$120$~{\rm deg}^2$ to quantify the number, extent, and distribution of HH objects in Serpens and the Aquila Rift. We incorporated \textit{Gaia} astrometry and dust maps from \citet{Edenhofer24} and \citet{Bayestar19} to better understand the full extent of the complex in three-dimensional space, as well as the interplay between gas, stars, and sites of active star formation. We see remarkable agreement, typically within a few percent, between progenitor source distances based on \textit{Gaia} data and distances from dust, calibrating the use of dust maps to anchor the distances to HH objects onto the region's substructures. This facilitates the identification of inter-cloud relationships, paving the way for the exploration of the star formation histories and mechanisms across this vast region. Our key findings are as follows:

\begin{enumerate}
    \item  We found 88 new HH objects, more than tripling the total known in the region.  Nearly all are located in or near fragmented cloud structures in ``translucent" ($1 < A_V < 5$) regions.  All clouds that met these criteria were targeted; thus it is unlikely that surveys of similar sensitivity will find many more HH objects in the Rift.
    \item Based on proximity and alignment with the HH objects, we identified candidate progenitor sources associated with the outflows found in our survey (Table \ref{tbl:progens}). Using cross-matches with the AllWISE and Spitzer Cores2Disks catalogs (Table \ref{tbl:progenitors_class}), the candidate progenitors were characterized-- confirming young star classifications in 55 of the 66 cases. Most of the 11 exceptions are also likley to be YSOs based upon previous studies.
    \item By combining HH object locations with dust maps from the literature, we determined that HH objects outline five spatially distinct regions of active star formation in the Serpens-Aquila region: (1) the Serpens Molecular Cloud complex, which contains Serpens Main, Northeast, and South/W40; (2) LDN~673, which contains the stellar population SCYA-59; (3) Serpens Near, which is much closer at $d\sim$250~pc; and (4-5) two components in the Aquila Rift, with a nearer component at $d\sim$600~pc in the west and a more distant component at $d\sim$700~pc in the east. 
    \item The Serpens Near cloud has velocities consistent with the edge of the local bubble and with the young component Scutum North association (ScuN-Y), which is directly adjacent to it. The active star formation indicated by the HH objects in Serpens Near may therefore mark a continuation of that previous stellar generation. 
    \item In the Serpens Molecular Cloud and Western Aquila Rift, HH objects trace the edges of low-density cavities that enclose a stellar population, as the Serpens Cloud forms the rear edge of a cavity around the MWC 297 cluster, and LDN~617 and adjoining clouds surround a component of group 332 from \citet{Prisinzano22}. This is consistent with feedback-driven triggered star formation, with shell gas that currently hosts HH objects being compressed by stellar feedback. 
\end{enumerate}

While we began this work with a sense of the breadth of the lateral extent of the region, we did not appreciate the much greater {\it depth} encompassed by the Rift and the complexity of the dense dust, stars, voids, gas, and HH objects distributed throughout this region. Fig.~\ref{fig:dust3d} illustrates how deceptive the appearance of the uniform cloud complex suggested by Fig.~\ref{fig:hh_loc} can be: while the dust extinction maps of \citet{Dobashi11} provide the illusion of a coherent series of related structures, in reality the Aquila Rift and Serpens extend many hundreds of parsecs into the Galaxy. The region is actually a multitude of stellar associations, cavities, and dense gas and dust structures in which pre-stellar cores \citep{2015A&A...584A..91K} and protostars are embedded. The presence of the latter is betrayed by the HH objects breaking out of the molecular cores.

Many of the HH objects discovered in Serpens Near and the Aquila Rift lack any clear connection to known stellar progenitors in \textit{Gaia}, highlighting how HH objects trace active star formation even while their parent stars remain unidentified. In this paper, we have used visible-light observations to identify dozens of new HH objects that have broken out of their natal cores, many of them traveling up to several parsecs from their progenitors-- exemplifying their utility as critical signposts for nodes of active star formation that can be targeted in future studies of this rich complex. Observatories such as the Roman Space Telescope will open embedded environments to highly sensitive, infrared astrometric measurements, facilitating the distances and motions of the progenitors to be determined, and supporting far deeper analyses of the region's structure and star formation history.




\begin{acknowledgments}
We are grateful to the National Science Foundation for support through awards AST-2206443 and AST-2206703 to TR and LP respectively. This research has made use of ``Aladin sky atlas" developed at CDS, Strasbourg Observatory, France.  This publication makes use of data products from the Wide-field Infrared Survey Explorer, which is a joint project of the University of California, Los Angeles, and the Jet Propulsion Laboratory/California Institute of Technology, funded by the National Aeronautics and Space Administration.  This research has made use of the NASA/IPAC Infrared Science Archive, which is funded by the National Aeronautics and Space Administration and operated by the California Institute of Technology.  
This work has made use of data from the European Space Agency (ESA) mission Gaia.  We thank B. Reipurth for the numbering of the HH objects and helpful comments. We also wish to thank Cerro Tololo Interamerican Observatory and its excellent support staff.  The figures in this paper were created with the help of the NOIRLab/IPAC/ESA/STScI/CfA FITS Liberator. The Dunlap Institute is funded through an endowment established by the David Dunlap family and the University of Toronto.

This publication has been made possible by the participation of over 1600 volunteers in our Zooniverse project, who completed $\sim$680,000 classifications of more than 45,000 cutouts. The individual contributors are acknowledged at: \url{https://www.zooniverse.org/projects/astrotrav/baby-star-search/about/results}.  We wish to thank Coleman Krawczyk and Joe Jackson for processing our Zooniverse data through their Caesar data aggregation code.

\end{acknowledgments}





%
\facilities{Blanco, WISE, Gaia, IRSA}

\software{}





\bibliography{rector, intro, shuping}{}
\bibliographystyle{aasjournal}
 




\end{document}